\documentclass{article}
\usepackage{cite}
\usepackage{graphicx}
\usepackage{latexsym}
\usepackage{amsmath}
\usepackage{amssymb}
\usepackage{amsfonts}
\usepackage{amsxtra}

\allowdisplaybreaks
\renewcommand{\L}{\mathcal{L}}
\newcommand{\I}{\mathcal{I}}
\renewcommand{\O}{\mathcal{O}}
\newcommand{\lambdat}{\widetilde{\lambda}}
\title{Steady states of the parametric rotator and pendulum} \author{Antonio O.\ Bouzas
  \thanks{E-mail: abouzas@fis.mda.cinvestav.mx}\\\small Departamento de
  F\'{\i}sica Aplicada, CINVESTAV-IPN \\\small Carretera Antigua a
  Progreso Km.\ 6, Apdo.\ Postal 73 ``Cordemex''\\\small M\'erida
  97310, Yucat\'an, M\'exico} 
\date{}
\begin{document}
\maketitle
\begin{abstract}
  We discuss several steady-state rotation and oscillation modes of
  the planar parametric rotator and pendulum with damping.  We
  consider a general elliptic trajectory of the suspension point for
  both rotator and pendulum, for the latter at an arbitrary angle with
  gravity, with linear and circular trajectories as particular cases.
  We treat the damped, non-linear equation of motion of the parametric
  rotator and pendulum perturbatively for small parametric excitation
  and damping, although our perturbative approach can be extended to
  other regimes as well.  Our treatment involves only ordinary
  second-order differential equations with constant coefficients, and
  provides numerically accurate perturbative solutions in terms of
  elementary functions.  Some of the steady-state rotation and
  oscillation modes studied here have not been discussed in the
  previous literature.  Other well-known ones, such as parametric
  resonance and the inverted pendulum, are extended to elliptic
  parametric excitation tilted with respect to gravity.  The results
  presented here should be accessible to advanced undergraduates,
  and of interest to graduate students and specialists in
  the field of non-linear mechanics.
\end{abstract}

\section{Introduction}
\label{sec:intro}

A physical system is said to be parametrically excited when some of
its parameters vary periodically with time.  One of the simplest
examples of a parametrically excited system consists of a rigid
rotator (or, in the presence of gravity, a physical pendulum) whose
suspension point follows a periodic trajectory, with both rotator and
suspension point restricted to move on a plane.  The basic modes of
motion of that system are rotation and oscillation.  Depending on the
type of parametric excitation, its amplitude, and the strength of
damping, many different types of rotation and oscillation modes may be
encountered in its spectrum of steady states.  Furthermore, those
basic modes combine to form complex motions in which rotation and
oscillation alternate in regular or irregular patterns, including
chaotic motion.  A qualitative description of a variety of different
modes of motion in a parametric pendulum, and some further references,
is given in the introduction of \cite{but01}.

The parametric rotator does not seem to have received much attention
in the literature, despite the fact that its rich non-linear dynamics
(see, for example, \cite{but02}) is interesting for both research and
education purposes.  The parametric pendulum is discussed in the
literature in relation to its two most remarkable dynamical phenomena.
One of them is parametric resonance, that occurs in systems
possessing normal modes of oscillation when the period of the
parametric excitation is an integer or half-integer multiple of the
period of a normal mode.  Parametric resonance in a pendulum with
vertically or horizontally moving suspension point is discussed in the
textbooks \cite{lan76a,arn78a}, and in \cite{yor78,cas80,gor82,cur95}.
Parametric resonance has also been discussed in the context of the
elastic or spring pendulum \cite{cay77,fal79,ani93,yan10}, a linear
oscillator \cite{but04,but05}, elastic strings \cite{fal79,row04}, 
electronic circuits \cite{fal79,ber02}, and many other systems.

The other striking dynamical aspect of the parametric pendulum is the
so-called ``inverted pendulum.''  A vertical parametric excitation,
with sufficiently large amplitude and frequency, causes the upright,
usually unstable, equilibrium position of the pendulum to become
stable. The dynamical stabilisation involved in this phenomenon is of
interest by itself, and also because it constitutes a mechanical
analogy for similar mechanisms in more complex systems
\cite{fri82,mic85}.  A detailed study of the inverted pendulum was
given by Kapitza \cite{kap65}.  Textbook treatments are given in
\cite{lan76,arn78}.  In the more recent literature a significant
number of studies of the inverted pendulum exist, both experimental
\cite{mic85,nes67,smi92,fen98} and theoretical
\cite{phe65,bli65,kal70,arn78,yor78,fri82,pip87,bla92,ach95,gra97,but01,but02,mat04}
(see also the references cited 
in those papers).  Most of those theoretical works deal with the
non-linear equation of motion of the parametric pendulum by
linearising it in the small-oscillation approximation
\cite{phe65,bli65,arn78,bla92,gra97} in which it reduces to a Mathieu
equation (see, e.g., \cite{rub96} and references therein), or within
the framework of the effective potential by separation of slow and
fast variables \cite{lan76,but01,but02}, or by considering
non-sinusoidal motion of the suspension point
\cite{kal70,yor78,pip87}.  In most cases damping is justifiably
neglected.  All of those theoretical approaches lead to important
insights into the dynamics of the inverted pendulum, and to
quantitative results such as, for example, the determination of its
region of stability.

In this paper we adopt a complementary point of view.  We look for
steady-state solutions to the damped, non-linear equation of motion of
the parametric rotator and pendulum.  Our approach is based on the
observation that the basic steady-state rotation and oscillation modes
are mathematically simple, in the sense that they can be expressed in
terms of elementary functions in some perturbative expansion.  For
such basic modes the non-linearity of the equation of motion manifests
itself in the multiplicity of different regimes, characterised by a
few dimensionless parameters, leading to different types of solution
and requiring different perturbative expansions.  Here, we restrict
ourselves to small parametric excitations and low to moderate damping
(both to be quantified below), and set up a perturbative expansion in
the small excitation amplitude.  In that regime the spectrum of basic
modes of motion is simpler than at larger amplitudes, though still
complex, but more interesting than in the limit of high friction.  The
cases of large parametric excitations and/or large friction could, in
principle, be handled by straightforward modifications of our
perturbative method.  Some of the steady-state rotation and
oscillation modes studied here have not been discussed in the previous
literature.  Other well-known ones, such as parametric resonance and
the inverted pendulum, are extended to elliptic parametric excitation
tilted with respect to gravity.

A pedagogical advantage of the approach adopted here is that it yields
numerically accurate approximate solutions to the equations of motion
of realistic non-linear systems like the parametric rotator and
pendulum by elementary mathematical methods. The only technical
requirements are the ability to perform power-series expansions and to
solve inhomogeneous ordinary second-order differential equations with
constant coefficients, which should be familiar from a course on
analytical mechanics.  Thus, we hope that the analysis in the
following sections may contribute to our understanding of non-linear
mechanics, and to make the latter more widely accessible to both
students and educators.

The paper is organised as follows.  In the next section we derive the
equation of motion for the rotator with general elliptic motion of its
suspension point, subject to gravity and damping, and establish our
notation and conventions.  In section \ref{sec:stead} we discuss the
two-dimensional rotator with small linear parametric excitation, in
the absence of gravity, explaining in detail the perturbative method
to solve the equation of motion for the steady-state of the system.
The same perturbative approach is applied in the subsequent sections.
The extension of the results of section \ref{sec:stead} to general
elliptic parametric excitation is carried out in section
\ref{sec:ellipt}.  The inclusion of gravity is considered in section
\ref{sec:gravell}, where we study rotation and oscillation modes of
the parametric pendulum elliptically excited at an arbitrary angle
with gravity.  Steady states related to parametric resonance are
discussed separately in section \ref{sec:parres}.  Finally, in section
\ref{sec:finrem} we give our summary and final remarks.

\section{The parametric rotator}
\label{sec:rotat}

We consider a two-dimensional rotator, that is, a rigid body free to
rotate about one of its points whose trajectory in space is fixed.  We
call the rotator symmetric if its suspension point is at its centre of
mass, otherwise asymmetric.  If a rotator is symmetric, a spatially
constant force field, such as an inertial or gravitational force,
cannot exert a torque on it.  To be specific, we will think of a
thin homogeneous bar of mass $M$ and length $L$ with its suspension
point at a distance $L_1$ from one of its ends and $L_2$ from the
other ($L_1+L_2=L$, $L_1>L_2$).  If we denote by $\vec{r}_0$ the
position of the suspension point, and by $\xi$ a coordinate along the
bar, $-L_2\leq \xi \leq L_1$, the points on the bar are parameterised
by $\vec{r}(\xi,t) = \xi \sin\theta \widehat{x} - \xi \cos\theta
\widehat{z} + \vec{r}_0$.  

It will be convenient to choose a coordinate system such that the
gravitational force points not in the direction $-\widehat{z}$, but
forming an angle $-\alpha$ with it.  The rotator's Lagrangian is then,
up to total derivatives,
\begin{equation}
  \label{eq:lag}
  \begin{aligned}
  \L &= \frac{1}{2} \int_{-L_2}^{L_1}d\xi \frac{M}{L}
  \dot{\vec{r}}(\xi,t)^{2} - \int_{-L_2}^{L_1}d\xi \frac{M}{L} g
  (\cos\alpha\, z_0 - \cos\alpha\cos\theta\,\xi + \sin\alpha\, x_0  
  +\sin\alpha \sin\theta\,\xi)\\
  &= \frac{1}{2} \I \dot{\theta}^2 + \frac{1}{2} M (L_1-L_2) \left(
    \sin\theta\, \ddot{x}_0 - \cos\theta\, \ddot{z}_0\right) +
  \frac{1}{2} M \dot{\vec{r}}_0^{\,2} + \frac{1}{2} Mg (L_1-L_2)
  \cos(\theta +
  \alpha)\\
  &\quad - Mg (\cos\alpha\, z_0 + \sin\alpha\, x_0)~,
  \end{aligned}
\end{equation}
where $\I = M/3(L_1^2+L_2^2-L_1L_2)$ is the moment of inertia with
respect to the suspension point, and where in the second equality we
dropped a total derivative. In what follows we always take into
account the effect on the rotator of friction, in the form of viscous
damping proportional to its angular velocity.  Adding such term to the
equation of motion for $\theta$ from the Lagrangian (\ref{eq:lag}) we
obtain
\begin{equation}
  \label{eq:eqaux}
  \I(\ddot{\theta} + \lambda \dot{\theta}) = \frac{1}{2} M(L_1-L_2)
  \left( \cos\theta\, \ddot{x}_0 + \sin\theta\, \ddot{z}_0\right) -
  \frac{1}{2} Mg(L_1-L_2) \sin(\theta+\alpha)~,
\end{equation}
$\lambda$ being the damping coefficient.  On the left-hand side of
(\ref{eq:eqaux}) the first term corresponds to the rotator's inertia
and the second one to damping due to friction.  On the right-hand side, the
first term describes the torque applied on the rotator by the inertial
forces due to the accelerated motion of its suspension point, and the
second one the torque due to gravity.  Notice that the right-hand side
vanishes for a symmetric rotator. 

We restrict ourselves to the case where the suspension point traces an
elliptic trajectory about the origin with a single frequency $\gamma$,
\begin{equation}
  \label{eq:ellips}
  x_0(t) = -r_0 \sin\epsilon \sin(\gamma t) ~,
\qquad
  z_0(t) = r_0 \cos(\gamma t)~,
\end{equation}
with $r_0$ the length of its semi-major axis, $r_0 \sin\epsilon$ that
of its semi-minor axis.  Notice that we have chosen the $z$ axis along
the major axis of the ellipse (\ref{eq:ellips}), which explains why we
took gravity along an arbitrary direction in (\ref{eq:lag}).
Substituting (\ref{eq:ellips}) in (\ref{eq:eqaux}) we obtain the
equation of motion for the rotator with an oscillating suspension
point.  It is convenient, however, to change the time variable in
(\ref{eq:eqaux}) to dimensionless time $\tau=\gamma t$, so as to
express the equation in terms of a minimal set of dimensionless
parameters.  After doing so, we obtain,
\begin{equation}
  \label{eq:eqgral}
  \frac{d^2\theta}{d\tau^2} + \lambdat
  \frac{d\theta}{d\tau} + \Delta \frac{x_0(\tau)}{r_0} \cos\theta +
  \Delta \frac{z_0(\tau)}{r_0} \sin\theta + \Gamma \sin(\theta+\alpha)
  =0~, 
\end{equation}
with  $\lambdat = \lambda/\gamma$ the dimensionless damping
coefficient and
\begin{equation}
  \label{eq:thinbar}
  \Delta = \frac{M(L_1^2-L_2^2)}{2\I} \frac{r_0}{L}~,
  \qquad
  \Gamma = \frac{M(L_1^2-L_2^2)}{2\I} \frac{g}{L\gamma^2}~.  
\end{equation}
Equation (\ref{eq:eqgral}) is generally valid for any rotator with its
suspension point oscillating as in (\ref{eq:ellips}), and
coefficients $\Delta$, $\Gamma$ depending on its geometry and mass
distribution.  For a symmetric rotator $\Delta=0=\Gamma$.  For the
homogeneous thin bar discussed above, $\Delta$ and $\Gamma$ are given by
(\ref{eq:thinbar}).

\section{Linearly excited parametric rotator}
\label{sec:stead}

We consider first the case in which the plane of the rotator is
horizontal, so that gravity does not play a role ($\Gamma=0$ in
(\ref{eq:eqgral})), and the motion of its suspension point is linear
($\epsilon=0$ in (\ref{eq:ellips})).  The equation of motion
(\ref{eq:eqgral}) then takes the form
\begin{equation}
  \label{eq:eq}
  \frac{d^2\theta}{d\tau^2} + \lambdat
  \frac{d\theta}{d\tau} + \frac{\Delta}{2} \sin(\theta+\tau) +
  \frac{\Delta}{2} \sin(\theta-\tau) =0~.
\end{equation}
Notice that in (\ref{eq:eq}) we can always choose the time origin so
that $\Delta>0$, and that the equation is invariant under
$\theta\rightarrow -\theta$.  We restrict ourselves to the regime of
small parametric excitations $0<\Delta\lesssim 1$.

\subsection{Basic solution}
\label{sec:motiv}

We look for steady-state solutions to (\ref{eq:eq}) with
$\theta(\tau)$ linearly rising with $\tau$, and possibly performing
small oscillations about that linear trajectory.  Thus, we are led to
propose a solution of the form
\begin{equation}
  \label{eq:thetaaux}
  \theta(\tau) = \omega\tau + \Theta_0 + f(\tau)\Delta~,
\quad
\omega \neq 0~,
\end{equation}
with $\Theta_0$ a constant phase.  It is physically reasonable to
include such constant term in $\theta(\tau)$ because, due to the
motion of the suspension point, the $z$ axis is a privileged direction
in space, making (\ref{eq:eq}) sensitive to the phase of $\theta$.
For instance, $\theta=n\pi$ with integer $n$ are equilibrium solutions
to (\ref{eq:eq}), but any other constant is not.  The last term in
(\ref{eq:thetaaux}) represents small oscillations with an amplitude of
$\O(\Delta)$, which requires the unknown function $f(\tau)$ to be
bounded for $0\leq\tau<\infty$.

If we substitute (\ref{eq:thetaaux}) in (\ref{eq:eq}) and let
$\Delta\rightarrow 0$, we find that $\omega\lambdat\rightarrow 0$.  We
must therefore assume that $\lambdat=\O(\Delta)$, which expresses the
physically reasonable condition that, for small parametric excitations
to cause steady-state rotations, friction must be correspondingly
small. (The relation between $\lambdat$ and $\Delta$ is made precise
below in equation (\ref{eq:lambda}).)  Thus, substituting
(\ref{eq:thetaaux}) in (\ref{eq:eq}) and neglecting terms of
$\O(\Delta^2)$ yields
\begin{equation}
  \label{eq:eqfirstaux}
  \begin{aligned}
    \frac{d^2f}{d\tau^2} = -\omega\frac{\lambdat}{\Delta} &-
    \frac{1}{2} \sin\left((\omega+1)\tau\right) \cos\Theta_0 -
    \frac{1}{2} \cos\left((\omega+1)\tau\right) \sin\Theta_0
    \\
    &\,- \frac{1}{2} \sin\left((\omega-1)\tau\right) \cos\Theta_0
    - \frac{1}{2} \cos\left((\omega-1)\tau\right) \sin\Theta_0~,
  \end{aligned}
\end{equation}
where $\lambdat/\Delta$ remains finite as $\Delta\rightarrow 0$.
Equation (\ref{eq:eqfirstaux}) is inconsistent with our assumptions,
since the constant term on its right-hand side leads to an unbounded
perturbation $f(\tau) = -1/2(\omega\lambdat/\Delta) \tau^2 + \cdots$,
growing quadratically with $\tau$.  The only way around this
inconsistency is to set $\omega=\pm1$, to obtain,
\begin{equation}
  \label{eq:eqfirst}
    \frac{d^2f}{d\tau^2} = \mp \frac{\lambdat}{\Delta} -
    \frac{1}{2} \sin(\Theta_0) \mp \frac{1}{2}
    \sin(2\tau\pm\Theta_0)~,
\qquad
\omega=\pm 1~.   
\end{equation}
Setting now $\sin(\Theta_0) = \mp2\lambdat/\Delta$, the
$\tau$-independent terms on the right-hand side of (\ref{eq:eqfirst}) vanish, 
and the equation admits a bounded solution $f(\tau)$.  We have then a
solution to (\ref{eq:eq}) of the form $\theta(\tau) =\pm
\theta_{0}(\tau) + \O(\Delta)$ with,
\begin{equation}
  \label{eq:zeroth}
  \theta_{0}(\tau) = \tau + \Theta_0~,
  \qquad
  \Theta_0 = -\arcsin(2\lambdat/\Delta) + 2n\pi~. 
\end{equation}
This result shows that for small $\Delta$ there cannot be
parametrically excited steady-state rotations unless
\begin{equation}
  \label{eq:lambda}
  \lambdat \leq \Delta/2~.
\end{equation}
We notice also that the steady-state phase $\Theta_0$ is completely
determined by the non-linear dynamics and the damping, without
reference to the initial conditions which, for times $\tau \gg
1/\lambdat$, are completely erased by the damping interactions.  We
conclude that, in physical units of time, at small $\Delta$ and
$\lambdat$ satisfying (\ref{eq:lambda}), steady-state rotations occur
only with the same angular frequency $\gamma$ as the parametric
excitation, $\theta(t) = \pm (\gamma t + \Theta_0) +
\mathcal{O}(\Delta)$, either clockwise or
counter-clockwise.

\subsection{Perturbation theory: steady-state rotation}
\label{sec:pert}

We now develop the considerations of the previous section into a
systematic perturbative approach.  We assume a solution to
(\ref{eq:eq}) of the form,
\begin{equation}
  \label{eq:solpert}
 \theta(\tau) = \theta_{0}(\tau) + \sum_{n=1}^\infty
 \frac{\Delta^n}{n!} (f_n(\tau) + \Theta_n)~,
\end{equation}
with $\theta_{0}(\tau)$ given by (\ref{eq:zeroth}).  We consider
only positive-frequency solutions $\theta(\tau)$ since, as noticed
above, negative-frequency ones are just given by $-\theta(\tau)$. The
perturbations $f_n(\tau)$ in (\ref{eq:solpert}) are assumed to be
bounded in $0\leq\tau<\infty$, uniformly with respect to $n$.  We have
explicitly separated from $f_n$ a constant term $\Theta_n$, so we
further require that $f_n(\tau)$ must not contain constant terms.  The
constants $\Theta_n$ are perturbative corrections to the zeroth-order
phase $\Theta_0$ in (\ref{eq:zeroth}).

Replacing (\ref{eq:solpert}) in (\ref{eq:eq}) and expanding to first
order in $\Delta$ leads to equation (\ref{eq:eqfirst}) (with upper signs).
With $\Theta_0$ given by (\ref{eq:zeroth}), and the condition that
$f_1(\tau)$ must not contain constant terms, we get a unique
solution to (\ref{eq:eqfirst})
\begin{equation}
  \label{eq:f1}
  f_1(\tau) = \frac{1}{8} \sin(2\tau+\Theta_0)~.
\end{equation}
In order to have a complete first-order solution we still need to find
the constant $\Theta_1$, which is fixed by requiring consistency of
the second-order equation.

To second order in $\Delta$ equation (\ref{eq:eq}) yields,
\begin{equation*}
  \frac{1}{2}\Delta^2 \frac{d^2f_2}{d\tau^2} + \lambdat\Delta
  \frac{df_1}{d\tau} + \frac{1}{2} \Delta^2 \cos(2\tau+\Theta_0) 
  (f_1(\tau)+\Theta_1) + \frac{1}{2} \Delta^2 \cos(\Theta_0)
  (f_1(\tau)+\Theta_1) =0~. 
\end{equation*}
Substituting $f_1$ in this equation we get, after some rearrangements,  
\begin{equation}
    \label{eq:eqsecndaux}
\frac{d^2f_2}{d\tau^2} = \cos(\Theta_0) \Theta_1 - \frac{3}{16}
\sin(2\tau) + \frac{1}{16} \sin(2\tau+2\Theta_0) - \frac{1}{16}
\sin(4\tau+2\Theta_0) ~.
\end{equation}
The first term on the right-hand side is a constant that must vanish for $f_2$
to be bounded, therefore
\begin{equation}
  \label{eq:theta1}
  \Theta_1 =0~.
\end{equation}
We can now find $f_2$ by twice integrating (\ref{eq:eqsecndaux}) with
respect to $\tau$.  The requirement that $f_2$ should not contain
zero-frequency terms leads to a unique solution,
\begin{equation}
  \label{eq:f2}
  f_2(\tau) = \frac{1}{2^8} \sin(4\tau+2\Theta_0) + \frac{3}{2^6}
  \sin(2\tau) - \frac{1}{2^6} \sin(2\tau + 2\Theta_0)~.
\end{equation}
With this result the solution to (\ref{eq:eq}) to $\O(\Delta^2)$ is
determined up to the constant $\Theta_2$, which results from requiring
consistency of the third-order equation.

We proceed further to consider the terms of $\O(\Delta^3)$ in equation
(\ref{eq:eq}) with (\ref{eq:solpert}).  After substituting $f_1$ and
$f_2$ in the third-order equation, somewhat lengthy algebra yields
\begin{equation}
  \label{eq:eqthird}
  \begin{aligned}
    \frac{d^2f_3}{d\tau^2} &= \left(-\frac{3}{2} \cos(\Theta_0)
      \Theta_2 + \frac{15}{2^8} \sin(\Theta_0)\right) 
    -\frac{3}{2} \cos(2\tau+\Theta_0) \Theta_2 - \frac{27}{2^8}
    \sin(2\tau-\Theta_0) + \frac{87}{2^{10}} \sin(2\tau+\Theta_0) \\ 
    &\quad-\frac{3}{2^8} \sin(2\tau+3\Theta_0) -
    \frac{45}{2^{10}}\sin(4\tau+\Theta_0) + \frac{15}{2^{10}}
    \sin(4\tau+3\Theta_0) - \frac{9}{2^{10}} \sin(6\tau+3\Theta_0)~.
  \end{aligned}
\end{equation}
As before, boundedness of $f_3$ means that the constant term in
brackets must vanish, so that,
\begin{equation}
  \label{eq:theta2}
  \Theta_2 = \frac{5}{2^7} \tan(\Theta_0)~.
\end{equation}
Now equation (\ref{eq:eqthird}) can be integrated, taking into account
that $f_3(\tau)$ must not contain constant terms, to yield,
\begin{equation}
  \label{eq:f3}
  \begin{aligned}
  f_3(\tau) &= \frac{3}{8} \Theta_2 \cos(2\tau+\Theta_0) +
  \frac{27}{2^{10}} \sin(2\tau-\Theta_0) - \frac{87}{2^{12}}
  \sin(2\tau+\Theta_0) + \frac{3}{2^{10}} \sin(2\tau+3\Theta_0)\\
&+
  \frac{45}{2^{14}} \sin(4\tau+\Theta_0) - \frac{15}{2^{14}}
  \sin(4\tau + 3\Theta_0) + \frac{1}{2^{12}} \sin(6\tau + 3\Theta_0)~.
  \end{aligned}
\end{equation}
The third-order correction is completed with the equality
\begin{equation}
  \label{eq:theta3}
  \Theta_3 = \frac{15}{2^9} \sin(\Theta_0)~,
\end{equation}
required for consistency of the fourth-order equation.

The above results constitute a perturbative solution to (\ref{eq:eq})
as a series expansion in $\Delta$, valid in the regime
$0<\Delta\lesssim 1$ and when the condition (\ref{eq:lambda}) holds.
The zeroth-order solution is (\ref{eq:zeroth}).  The first, second and
third order solutions are given by,
\begin{equation}
  \label{eq:firstsecondthird}
  \theta_1(\tau) = \theta_0(\tau)+ f_1(\tau) \Delta~,
  \quad
  \theta_2(\tau) = \theta_1(\tau)+ (\Theta_2+f_2(\tau))
  \frac{\Delta^2}{2}~, 
  \quad
  \theta_3(\tau) = \theta_2(\tau)+ (\Theta_3+f_3(\tau))
  \frac{\Delta^3}{6}~. 
\end{equation}
Several remarks about this perturbative solution are in order.  (a) In
the perturbative expansion (\ref{eq:solpert}) we could have included
corrections to the angular frequency, with a first term in
(\ref{eq:solpert}) of the form $(1+\omega_1 \Delta + \omega_2
\Delta^2/2 + \cdots) \tau$ instead of just $\tau$. All the corrections
$\omega_n$, however, turn out to vanish.  (b) The results
(\ref{eq:f1}), (\ref{eq:f2}), (\ref{eq:f3}), for $f_{1,2,3}$ suggest
that the perturbative expansion parameter is actually $\Delta/8$
rather than $\Delta$, so the domain of applicability of perturbation
theory should be larger than expected. This is confirmed below in
section \ref{sec:numlin} on numerical results.  (c) $\Theta_0$ is
determined only up to $2n\pi$.  The additional $n$ turns typically
build up during the transient state.  Our perturbative solutions hold
valid only for times $\tau\gg 1/\lambdat$ after all transients have
died off, and depend on $\lambdat$ only through the phase $\Theta_0$.
(d) The perturbations $f_n(\tau)$ contain only even frequencies (in
physical units, only even multiples of $\gamma$). Thus,
$\theta(\tau)-\tau$ is periodic with period $\pi$.

These results are numerically verified below in section
\ref{sec:numlin}, where they are compared with numerical solutions of
(\ref{eq:eq}) and found to be in tight agreement with them, each order
significantly improving on the previous one.  Those verifications
justify our perturbative procedure, together with the implicit
hypothesis that the dependence of $\theta(\tau)$ on $\Delta$ is
analytic at $\Delta=0$.

\subsection{Steady-state oscillations}
\label{sec:oscilin}

When condition (\ref{eq:lambda}) is not satisfied the steady-state
rotation described in the previous section does not exist.  In that
case, the rotator may reach the equilibrium solutions $\theta =
n\pi$.  There is another type of steady-state solution that is
possible in that case, however, describing oscillations about
$(2n+1)\pi/2$.    To obtain those solutions we assume they have the
form,
\begin{equation}
  \label{eq:pertzero}
 \theta(\tau) = \Theta_{0} + \sum_{n=1}^\infty
 \frac{\Delta^n}{n!} (f_n(\tau) + \Theta_n)~.
\end{equation}
Substituting this form into the equation of motion (\ref{eq:eq}) leads
to perturbative equations for $f_n$ that can be solved by the same
method as in section \ref{sec:pert}.  We will omit the details of the
procedure for brevity, commenting only on the form of the solution.

The condition that the perturbation $f_2(\tau)$ be bounded leads to
$\sin(2\Theta_0)=0$, whereas the equations at third, fourth and fifth
order yield $\Theta_1 =\Theta_2 =\Theta_3 = 0$.  If we set $\Theta_0 =
k\pi$, for some integer $k$, we recover the equilibrium solutions and
all corrections $\Theta_n$, $f_n$ must vanish because $\Theta_0$ is
itself an exact solution to the equation of motion.  Another possible
choice is $\Theta_0=(2k+1)\pi/2$, with $k$ an integer.  In that
case the solution is non trivial.  At first order we obtain,
\begin{equation}
  \label{eq:oscilin-first}
  \theta_1(\tau) = \Theta_0 + \Delta \sin(\Theta_0) \cos(\tau)~.
\end{equation}
The equation for $f_2$ yields the second-order solution,
\begin{equation}
  \label{eq:oscilin-scnd}
  \theta_2(\tau) = \theta_1(\tau) - \Delta\lambdat \sin(\Theta_0)
  \sin(\tau) ~,
\end{equation}
and at third order we find
\begin{equation}
  \label{eq:oscilin-thrd}
\theta_3(\tau) = \theta_2(\tau) -  \sin(\Theta_0)
\left(\frac{9}{24} \Delta^3 \cos(\tau) +  \Delta\lambdat^2
  \cos(\tau) + \frac{1}{72} \Delta^3\cos(3\tau)\right) ~.
\end{equation}
The perturbative method does not yield information on the domain of
validity of these solutions in parameter space.  Numerical analysis of
(\ref{eq:eq}) suggests that, in the perturbative regime $\Delta$,
$\lambdat \lesssim 1$, the steady state described by
(\ref{eq:oscilin-first})--(\ref{eq:oscilin-thrd}) exists only for
$\Delta\ll \lambdat$.

\subsection{Order-of-magnitude estimates}
\label{sec:damp}

The quantitative meaning of the parameter $\Delta$ in (\ref{eq:eq}) is
apparent from its definition (\ref{eq:thinbar}) for a homogeneous thin
bar, or its correspondingly modified expressions for other rigid
bodies.  Thus, for example, for a homogeneous thin bar with $L_1=5.4$
cm, $L_2=4.6$ cm and excitation amplitude $r_0=0.5$ cm we have
$\Delta=0.016$.  For a larger, more asymmetric thin bar with $L_1=13$
cm, $L_2=7$ cm and amplitude $r_0=2$ cm we obtain $\Delta = 0.094
\simeq 0.1$. By reducing the asymmetry or the amplitude we can get
values of $\Delta$ smaller by several orders of magnitude.  With
larger asymmetries and amplitudes $\Delta$ can be made larger,
$\Delta\gtrsim 1$.  For values $\Delta>2$ the parametric-excitation
amplitude must be larger than the length of the rotator itself, so
rather than parametrically excited, the rotator should in that case
more properly be referred to as orbitally excited.

In order to obtain some quantitative understanding of the parameter
$\lambdat$ in (\ref{eq:eq}), we consider the steady-state rotations
described in section \ref{sec:pert}.  Once the steady state has been
reached, we may cease the parametric excitation and find out how many
turns the rotator completes before it stops moving.  Since for a
steady-state rotation of the form (\ref{eq:solpert}) the
(dimensionless) angular velocity is $1+\mathcal{O}(\Delta)$, the
number of turns sought for must be approximately $1/(2\pi \lambdat)$
independently of $\Delta$.  Indeed, after the motion of the suspension
point has stopped, we find by numerically solving (\ref{eq:eq}) with
$\Delta=0$ that the rotator performs slightly more than 15 turns if
$\lambdat=10^{-2}$, about 150 turns if $\lambdat=10^{-3}$, and about
1500 turns for $\lambdat=10^{-4}$.  From our everyday experience we
know that the first case corresponds to the level of damping usually
encountered in common household appliances such as fans or mixers.
The second case corresponds to a much lower level of damping, found in
more special mechanical devices such as some microprocessor fans.  In
all three cases we find that the perturbative results of section
\ref{sec:pert} for the steady state agree with exact numerical
solutions to (\ref{eq:eq}) with high accuracy, as discussed in the
following section.  For much lower damping, say $\lambdat\lesssim
10^{-6}$, which can be achieved in laboratory experiments, the steady
state takes a long time to be reached.  In those cases, it may be
appropriate to modify our perturbative expansion by setting
$\lambdat=\mathcal{O}(\Delta^n)$ with $n>1$, or just solve
(\ref{eq:eq}) numerically for the entire time evolution, including the
long transient state.

\subsection{Numerical results}
\label{sec:numlin}

We now turn to comparing numerical solutions to (\ref{eq:eq}) with the
perturbative ones found in the previous sections.  We numerically
solved (\ref{eq:eq}) with standard commercial software \cite{wolf}
over an interval $0\leq\tau\leq \tau_\textmd{max}$ with
$\tau_\textmd{max}\gg 1/\lambdat$, with appropriate initial
conditions.  The type of steady state reached by the system is
insensitive to $\theta(0)$ in most cases, but may be sensitive
to $d\theta/d\tau(0)$ depending on the values of $\Delta$ and
$\lambdat$. 

\begin{figure}
  \centering
\begin{picture}(460,300)(0,0)
 \put(0,150){\scalebox{0.75}{\includegraphics{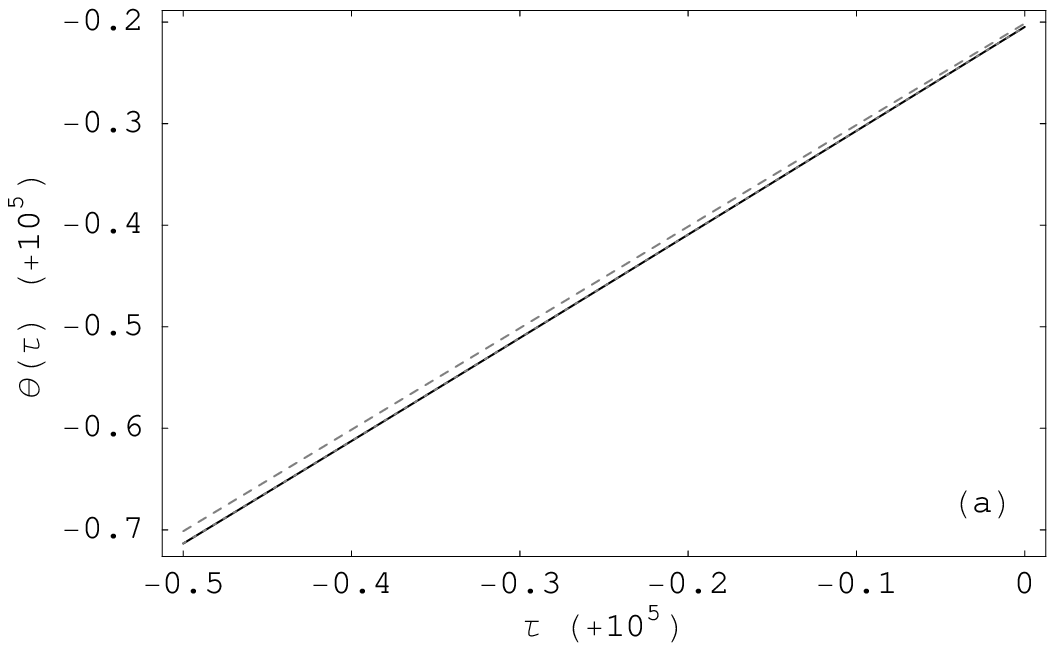}}}  
 \put(240,150){\scalebox{0.7395}{\includegraphics{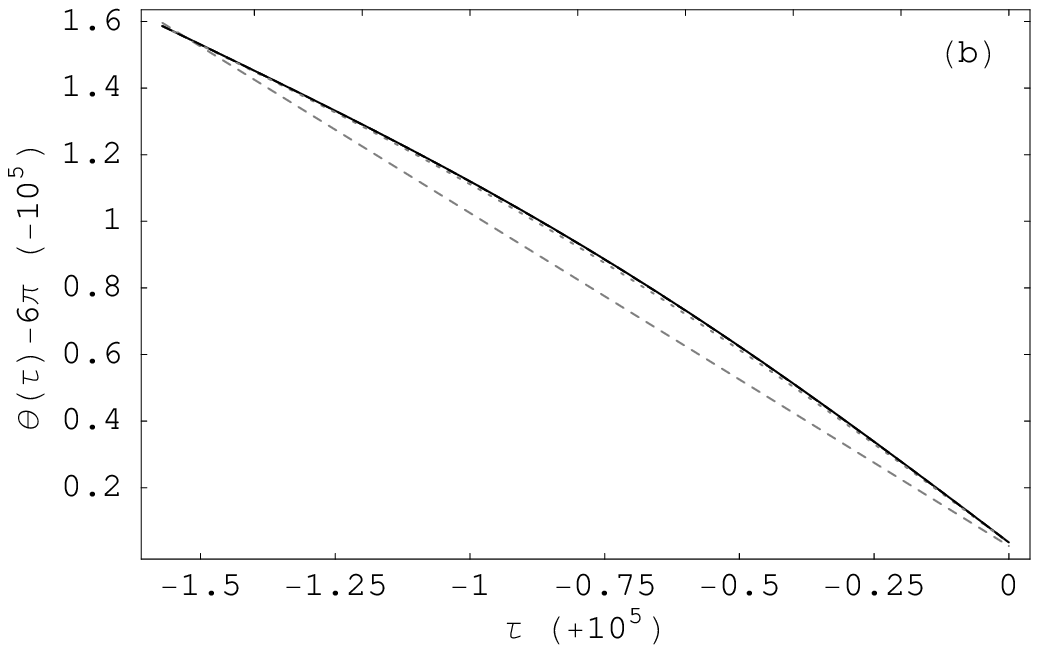}}}  
 \put(0,0){\scalebox{0.75}{\includegraphics{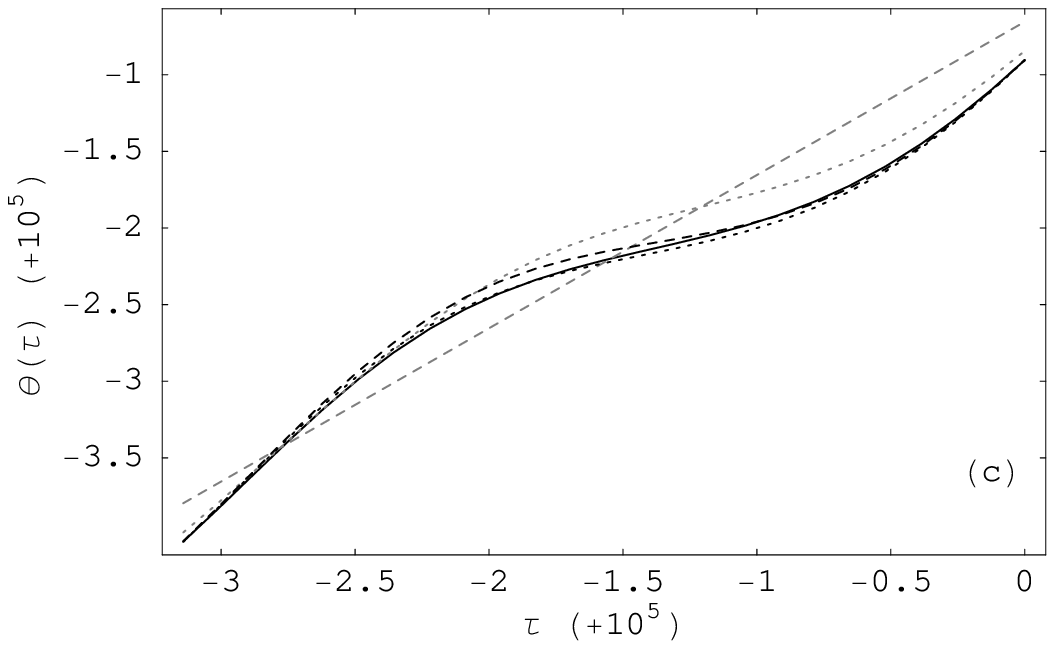}}}  
 \put(237,1){\scalebox{0.745}{\includegraphics{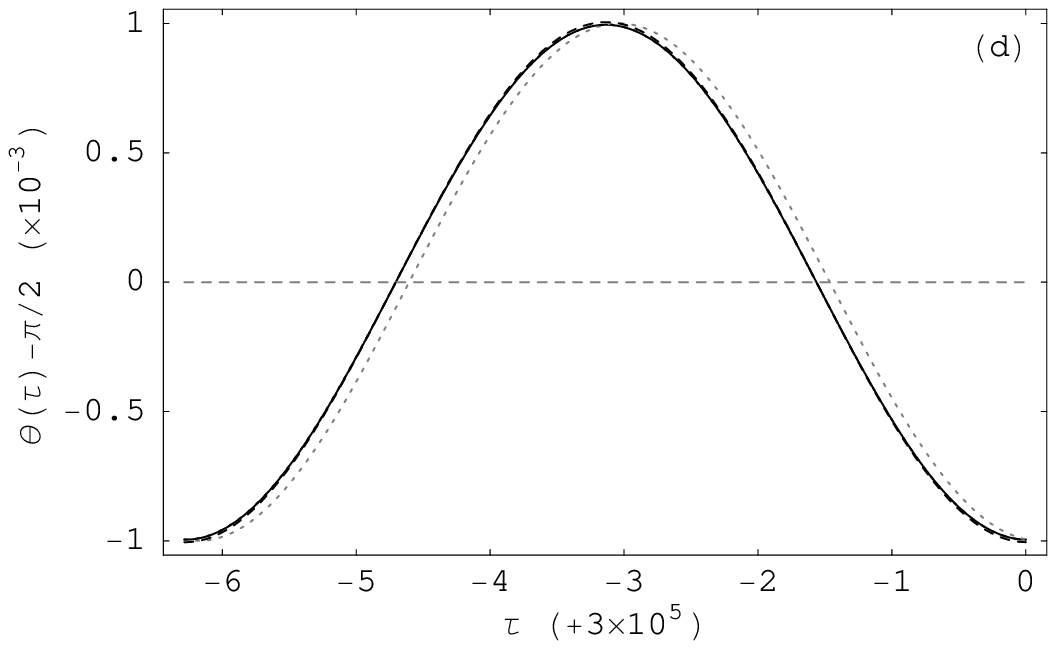}}}  
\end{picture}
  \caption{Numerical solution (black solid line) to equation
    (\ref{eq:eq}), compared with perturbative solutions at zeroth
    (grey dashed line), first (grey dotted line), second (black dashed
    line) and third (black dotted line) orders as given in section
    \ref{sec:stead}.  Steady-state rotations (see section
    \ref{sec:pert}: (a) $\Delta=0.1$, $\lambdat=0.01$, $\theta(0)=0$,
$d\theta/d\tau(0)=0.7$; 
(b) $\Delta=0.8$, $\lambdat=0.01$, $\theta(0)=0$, $d\theta/d\tau(0)=0.001$;
(c) $\Delta=2.3$, $\lambdat=0.7$, $\theta(0)=-0.1$,
$d\theta/d\tau(0)=1$. 
Steady-state oscillations (see section \ref{sec:oscilin}):
(d) $\Delta=0.001$, $\lambdat=0.1$, $\theta(0)=\pi/2$,
$d\theta/d\tau(0)=0$.}
  \label{fig:fig1}
\end{figure}

In figure \ref{fig:fig1} we show steady-state rotations for three
different values of $\Delta$ and $\lambdat$.  With $\Delta=0.1$ and
$2\lambdat/\Delta=0.2$, as shown in figure \ref{fig:fig1} (a), the
first-order approximation $\theta_1(\tau)$ from
(\ref{eq:firstsecondthird}) already agrees almost exactly with the
numerical solution.  For these parameter values we numerically find
steady-state rotations for any value of $\theta(0)$, and for $0.64
\leq d\theta/d\tau(0) \leq 2.13$.  Outside of that range of initial
angular velocities the steady states we obtain numerically are
the equilibria $\theta(\tau) = n\pi$.

In figure \ref{fig:fig1} (b) we set $\Delta=0.8$ and
$2\lambdat/\Delta=0.025$.  The first-order approximation is very close
to the numerical solution, and the second-order one $\theta_2(\tau)$
is indistinguishable from it.  We numerically find steady-state
rotations for any $\theta(0)\neq 2n\pi$, and for all values of
$10^{-6} \lesssim d\theta/d\tau(0) \lesssim 10^3$ we tried.

Numerical analysis of (\ref{eq:eq}) shows that the steady-state
rotations described in section \ref{sec:pert} for $\Delta\lesssim 1$
exist also for larger values of $\Delta$, although the perturbative
expansion discussed there is not directly applicable when $\Delta\gg
1$. (In the latter case, an expansion in powers of $1/\Delta$ would be
appropriate.) Nevertheless, as discussed at the end of section
\ref{sec:pert}, the form of the expansion suggests that it may
converge even for $\Delta\gtrsim 1$.  This is shown in figure
\ref{fig:fig1} (c), where $\Delta=2.3$ and $2\lambdat/\Delta=0.61$.
As seen in the figure, the agreement between the numerical and the
perturbative solutions is very accurate at third order.  Also, each
successive perturbative order gives a significant improvement over the
previous one, signalling good convergence of perturbation theory even
at such large values of $\Delta$ (as long as damping is also
sufficiently large). In this case we also  find steady-state
rotations for any $\theta(0)\neq 2n\pi$, and for all values of
$d\theta/d\tau(0)$ over many orders of magnitude.

The steady-state rotations described in section \ref{sec:pert} and in
figures \ref{fig:fig1} (a), (b), (c), exist for large open sets in
parameter space, even in the perturbative regime $\Delta$,
$\lambdat\lesssim 1$, and are reached by the system for relatively
large open sets of initial conditions, as described above.  The
steady-state oscillations described in section \ref{sec:oscilin}, by
contrast, are found numerically only for $\Delta\ll \lambdat$.  Under
those conditions, the angular velocity quickly reaches its typical
steady-state magnitude and, as a result, the mean value of $\theta$
evolves very slowly towards its steady-state value and numerical
solutions are computationally expensive.  Also, unlike steady
rotations, these oscillating steady-states are not robust in the sense
that even small variations in the initial velocity may lead the
system away from them and towards an equilibrium solution.  An example
of steady-state oscillations about $\pi/2$ is shown in figure
\ref{fig:fig1} (d).  In that case the numerical result is well
described by the first-order solution (\ref{eq:oscilin-first}) and, in
the figure, indistinguishable from the second-order one
(\ref{eq:oscilin-scnd}).

\section{Elliptically excited parametric rotator}
\label{sec:ellipt}

The simplest implementation of a parametric rotator is probably the
well-known toy model described in \cite{wal06} as a ``rotator on a
notched stick.''  In that case the trajectory of the suspension is
well approximated by a highly eccentric ellipse.  The elliptical
excitation of the parametric rotator is of interest by itself, and
also because it continuously connect the cases of linear ($\epsilon=0$)
and circular ($\epsilon=\pm\pi/2$) excitations, that lead to rather
different types of steady states.  In this section we consider the
extension of the results of section \ref{sec:stead} to the case of
elliptic motion of the suspension point.

The equation of motion (\ref{eq:eqgral}), without gravity, takes the
form,
\begin{equation}
  \label{eq:eqell}
  \frac{d^2\theta}{d\tau^2} + \lambdat
  \frac{d\theta}{d\tau} + \frac{\Delta}{2}(1-\sin\epsilon)
  \sin(\theta+\tau) + \frac{\Delta}{2}
  (1+\sin\epsilon)\sin(\theta-\tau) =0~, 
\end{equation} 
where $-\pi/2\le \epsilon \le \pi/2$.  For $\epsilon=0$ the suspension
point oscillates along a line segment, and we recover (\ref{eq:eq}).
For $0<\epsilon<\pi/2$ the suspension point moves counterclockwise
along an elliptic orbit, which becomes circular at $\epsilon=\pi/2$.
Similarly, negative values $-\pi/2\le \epsilon < 0$ describe clockwise
motion of the suspension point.  In this paper we restrict ourselves
mostly to eccentric elliptic excitations, $|\epsilon|\lesssim
1/2$ that are close to the linear case considered in previous
sections.  Circular, or slightly-eccentric elliptic, motions of the
suspension point ($1\lesssim|\epsilon|\le \pi/2$) lead to
qualitatively different solutions that will be discussed elsewhere
\cite{bouxx}. 

Equation (\ref{eq:eqell}) is invariant under the transformation
$\theta\rightarrow -\theta$, $\epsilon\rightarrow -\epsilon$. Thus, if
$\theta(\tau,\epsilon)$ is a solution to (\ref{eq:eqell}) for some
$\epsilon$, then $-\theta(\tau,-\epsilon)$ is also a solution for the
same $\epsilon$.  There are no equilibrium solutions to
(\ref{eq:eqell}) when $\epsilon\neq0$.  When $\epsilon=0$, the linear
case, the equilibria are $\theta(\tau)=n\pi$, with integer $n$.


\subsection{Steady-state rotations}
\label{sec:ellrotat}

We now look for counter-clockwise steady-state rotations of the form
(\ref{eq:solpert}).  Let $\theta(\tau,\epsilon_0)$ be such a solution
for $\epsilon=\epsilon_0>0$.  Since both the rotator and suspension
point move in the same sense, we will say that
$\theta(\tau,\epsilon_0)$ describes a ``direct motion'' of the
rotator.  By symmetry, $-\theta(\tau,-\epsilon_0)$ is a solution of
(\ref{eq:eqell}) with $\epsilon=\epsilon_0$, describing a clockwise
motion of the rotator with a counter-clockwise moving suspension
point, what we call a ``contrarian motion'' of the rotator.  Similarly,
$\theta(\tau,-\epsilon_0)$ and $-\theta(\tau,\epsilon_0)$ are
solutions to (\ref{eq:eqell}) with $\epsilon=-\epsilon_0<0$,
describing contrarian and direct motion, respectively.

After substituting (\ref{eq:solpert}) in (\ref{eq:eqell}), a
straightforward application of the perturbative method of section
\ref{sec:stead} leads to the perturbative solution we seek.  At lowest
order we get
\begin{equation}
  \label{eq:zeroth-ell}
  \theta_0(\tau) = \tau + \Theta_0~,
  \qquad
  \Theta_0 = -\arcsin\left(\frac{2}{1+\sin\epsilon}
    \frac{\lambdat}{\Delta}\right) + 2n\pi~.
\end{equation}
Therefore, in this case steady-state rotations occur only if
\begin{equation}
  \label{eq:lambda-ell}
  \lambdat < \frac{\Delta}{2} (1+\sin\epsilon)~,
\end{equation}
which generalises condition (\ref{eq:lambda}) to elliptic excitations
of the rotator.  We see also that contrarian motions are allowed by
(\ref{eq:lambda-ell}), for small enough $\lambdat$, except for
$\epsilon=-\pi/2$.  Thus, for circular parametric excitation there can
be no contrarian motion.

At first order we get,
\begin{equation}
  \label{eq:first-ell}
  \theta_1(\tau) = \theta_0(\tau) + \Delta f_1(\tau)~,
  \qquad
  f_1(\tau) = \frac{1}{8} (1-\sin\epsilon) \sin(2\tau+\Theta_0)~.
\end{equation}
In particular, $\Theta_1=0$ as in the linear-excitation case of section
\ref{sec:stead}.  The second-order solution takes the form,
\begin{equation}
  \label{eq:second-ell}
  \begin{gathered}
  \theta_2(\tau) = \theta_1(\tau) + \frac{\Delta^2}{2} (\Theta_2 +
  f_2(\tau))~,   
  \qquad
  \Theta_2 = \frac{5}{2^7} (1-\sin\epsilon)^2 \tan(\Theta_0)~,\\
f_2(\tau) = \frac{3}{2^6} (\cos\epsilon)^2 \sin(2\tau) - \frac{1}{2^6}
(\cos\epsilon)^2 \sin(2\tau + 2\Theta_0) + \frac{1}{2^8}
(1-\sin\epsilon)^2 \sin(4\tau+2\Theta_0)~.
  \end{gathered}
\end{equation}
Finally, at third order we obtain,
  \begin{equation}
    \label{eq:third-ell}
    \begin{aligned}
  \theta_3(\tau) &= \theta_2(\tau) + \frac{\Delta^3}{3!} (\Theta_3 +
  f_3(\tau)) ~,
  \qquad
  \Theta_3 = \frac{15}{2^9} (1-\sin\epsilon) (\cos\epsilon)^2
  \sin(\Theta_0)~, \\
  f_3(\tau) &= \frac{3}{8} \Theta_2 (1-\sin\epsilon)
  \cos(2\tau+\Theta_0)
  + \frac{27}{2^{10}} (\cos\epsilon)^2 (1+\sin\epsilon)
  \sin(2\tau-\Theta_0)\\
  &+ \frac{3}{2^{12}} (29 (\sin\epsilon)^3 + 9(\sin\epsilon)^2 - 9
  \sin\epsilon -29) \sin(2\tau+\Theta_0)
  + \frac{3}{2^{10}} (\cos\epsilon)^2 (1+\sin\epsilon)
  \sin(2\tau+3\Theta_0)\\
  &+ \frac{45}{2^{14}} (\cos\epsilon)^2 (1-\sin\epsilon)
  \sin(4\tau+\Theta_0)
  - \frac{15}{2^{14}} (\cos\epsilon)^2 (1-\sin\epsilon) \sin(4\tau +
  3\Theta_0)\\
  &+ \frac{1}{2^{12}} (1-\sin\epsilon)^3 \sin(6\tau + 3\Theta_0)~.
    \end{aligned}
  \end{equation}
Some remarks on these expressions are in order. First, at
$\epsilon=0$ the perturbative solution described above reduces to
the one found in section \ref{sec:pert} for the case of linear
excitation.  Second, at $\epsilon=\pi/2$ the perturbative
corrections $\Theta_{2,3}$ and $f_{1,2,3}(\tau)$ vanish.  In fact,
all $\Theta_n$, $f_n(\tau)$, $n\geq 1$, must vanish in that case
because for circular excitation $\theta_0(\tau)$, as given by
(\ref{eq:zeroth-ell}), is an exact solution to (\ref{eq:eqell}).
Third, only even angular frequencies enter the perturbative
corrections.

\subsection{Steady-state oscillations}
\label{sec:osci-ell}

We now look for solutions to (\ref{eq:eqell}) of the form
(\ref{eq:pertzero}).  Assuming $\Delta$, $\lambdat\lesssim 1$, and
substituting (\ref{eq:pertzero}) in (\ref{eq:eqell}), leads us to a
perturbative analysis completely analogous to that of section
\ref{sec:oscilin}, so we omit the details for brevity.  From
(\ref{eq:pertzero}) we have the zeroth-order solution
\begin{equation}
  \label{eq:zeroth-osciell}
  \theta_0(\tau) = \Theta_0,
\end{equation}
with $\Theta_0$ undetermined at this order.  The first-order equation
yields, 
\begin{equation}
  \label{eq:first-osciell}
  f_{(g)1}(\tau) = 
  -\frac{1}{2} (1+\sin\epsilon)
  \sin(\tau-\Theta_0) + \frac{1}{2} (1-\sin\epsilon)
  \sin(\tau+\Theta_0)~,
\end{equation}
Consistency of the second-order equation requires $\cos\epsilon
\sin(2\Theta_0) = 0$, therefore,
\begin{equation}
  \label{eq:theta0-osciell}
  \Theta_0 = \frac{1}{2} k\pi~, 
   \qquad \text{if} \quad \cos\epsilon \neq0~,
\end{equation}
for some integer $k$.  In the case of circular excitation,
$\cos\epsilon=0$, $\Theta_0$ as well as all $\Theta_n$ remain
undetermined due to the invariance of (\ref{eq:eqell}) under a
one-parameter continuous symmetry.  The details of that particular
case will be discussed elsewhere \cite{bouxx}.  With $\Theta_0$ given
by (\ref{eq:theta0-osciell}), the equation for $f_2$ leads to,
\begin{equation}
  \label{eq:second-osciell}
  \begin{aligned}
  f_2(\tau) &= (\Theta_1 - \lambdat/\Delta) (1+\sin\epsilon)
  \cos(\tau-\Theta_0) + (\Theta_1 + \lambdat/\Delta) (1-\sin\epsilon)
  \cos(\tau+\Theta_0)\\
  &\quad-\frac{1}{16} (1+\sin\epsilon)^2 \sin(2\tau-2\Theta_0) +
  \frac{1}{16} (1-\sin\epsilon)^2 \sin(2\tau+2\Theta_0)~.
  \end{aligned}
\end{equation}
The constant $\Theta_1$, undetermined at this order, is fixed by the
consistency requirement for the third-order equation
\begin{equation}
  \label{eq:theta1-osciell}
  \Theta_1 = \frac{\lambdat}{\Delta}
  \frac{\sin\epsilon}{(\cos\epsilon)^2} \cos(2\Theta_0)~.
\end{equation}
With this value for $\Theta_1$ we obtain a bounded solution
$f_3(\tau)$ to the third-order equation.  We omit the expression for
$f_3(\tau)$, however, because it is too lengthy to transcribe here.
Finally, 
\begin{equation}
  \label{eq:theta2-osciell}
  \Theta_2 = 0~
\end{equation}
is required for consistency of the fourth-order equation.

For $\epsilon =0$ these solutions must satisfy (\ref{eq:eq}).  Thus,
setting $k=2n$ in (\ref{eq:theta0-osciell}) leads to $\Theta_0=n\pi$,
and we recover the equilibrium solutions of the linear case.  In
particular, all corrections $\Theta_i$, $f_i$, $i\geq 1$ must vanish,
since $\Theta_0$ is an exact solution to (\ref{eq:eq}) in this case.
It is easily seen that this indeed happens with the corrections
(\ref{eq:first-osciell}) and
(\ref{eq:second-osciell})--(\ref{eq:theta2-osciell}).  On the other
hand, by setting $k=2n+1$ in (\ref{eq:theta0-osciell})  we recover
the oscillating solutions discussed in section \ref{sec:oscilin}. 

\subsection{Numerical results}
\label{sec:numell}

\begin{figure}
  \centering
\begin{picture}(460,300)(0,0)
 \put(0,150){\scalebox{0.75}{\includegraphics{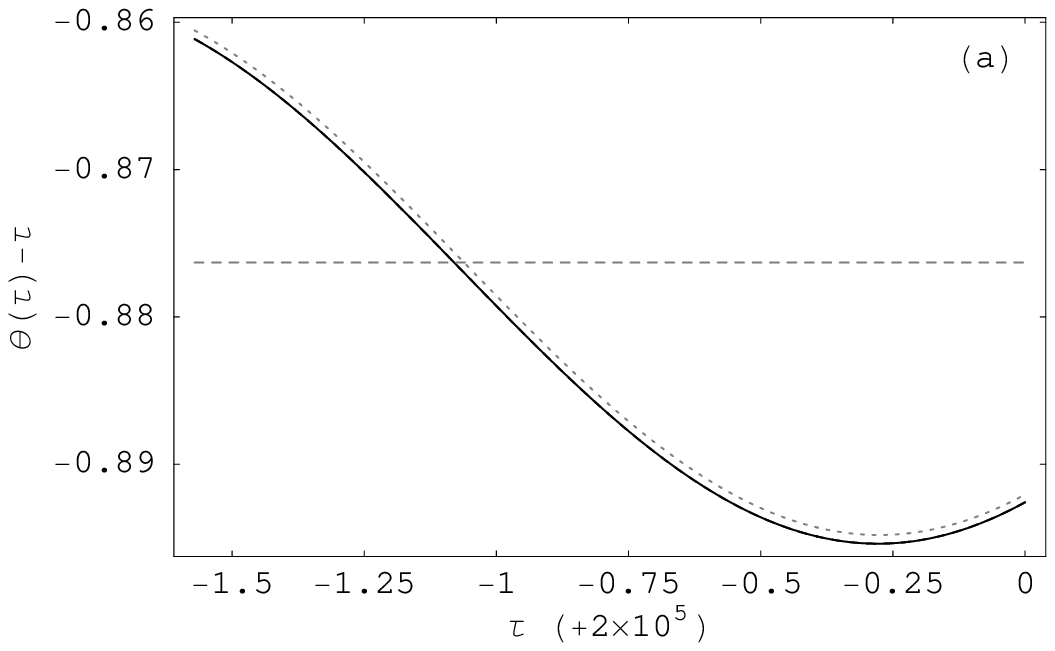}}}  
 \put(240,150.5){\scalebox{0.730}{\includegraphics{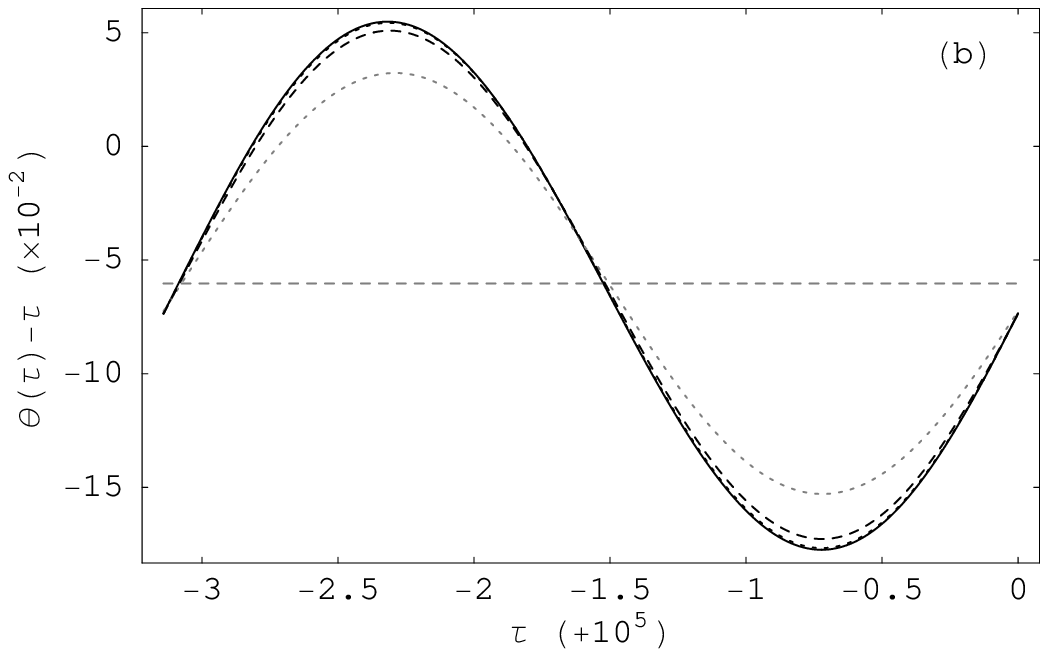}}}  
 \put(-6,0){\scalebox{0.77}{\includegraphics{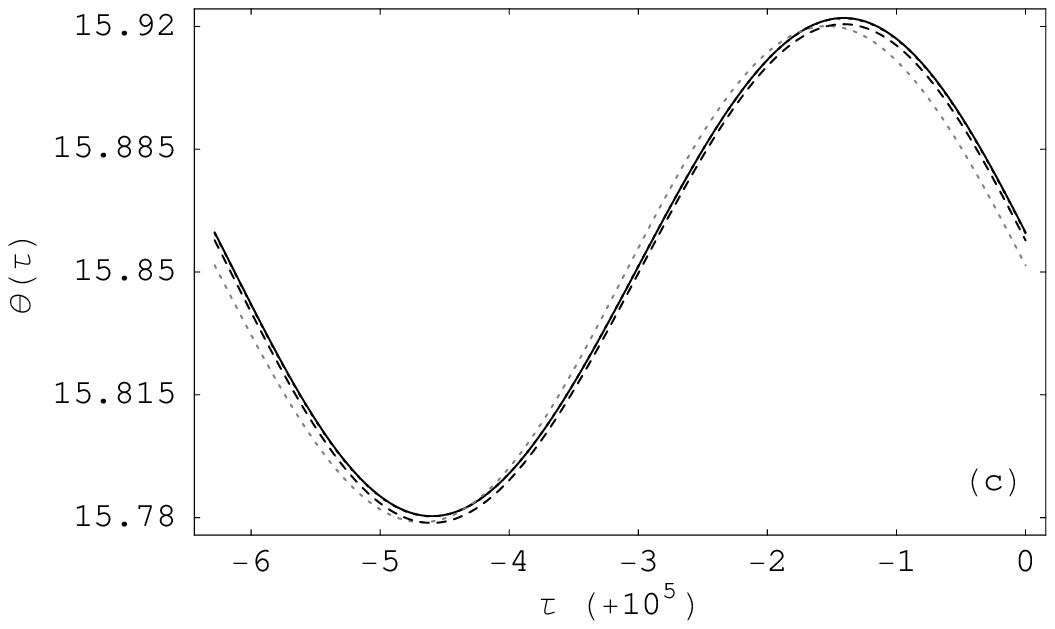}}}  
 \put(241,4){\scalebox{0.73}{\includegraphics{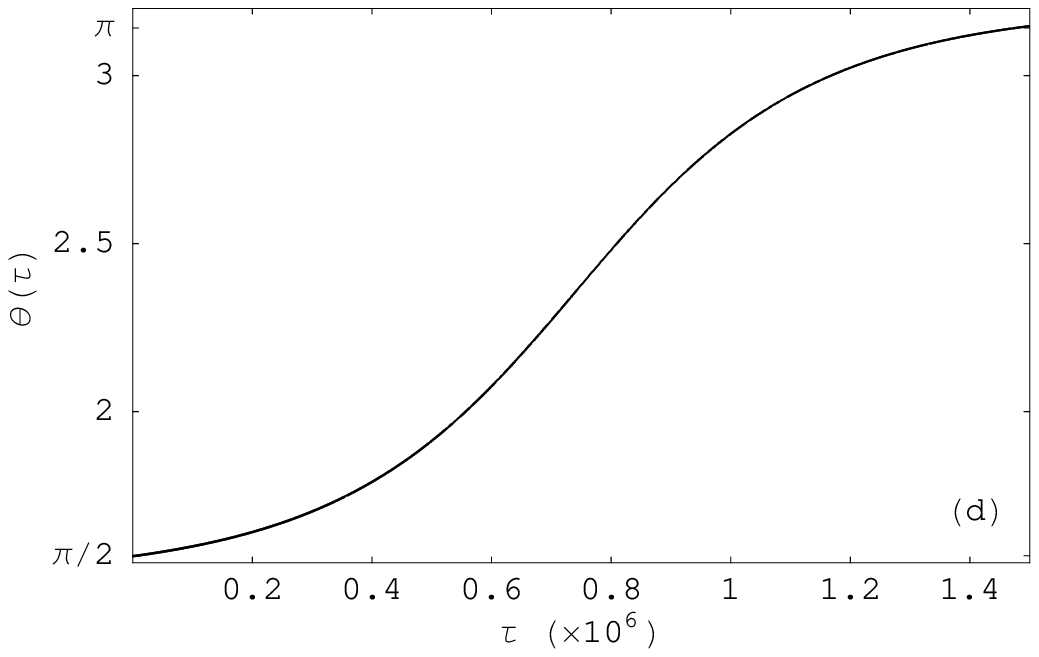}}}  
\end{picture}
  \caption{Numerical solution (black solid line) to equation
    (\ref{eq:eqell}), compared with perturbative solutions at zeroth
    (grey dashed line), first (grey dotted line), second (black dashed
    line) and third (black dotted line) orders as given in section
    \ref{sec:ellipt}.  Steady-state rotations (see section
    \ref{sec:ellrotat}):
(a) $\epsilon=-0.5$, $\Delta=0.1$, $\lambdat=0.02$,
$\theta(0)=0$,$d\theta/d\tau(0)=1$; 
(b) $\epsilon=\pi/8$, $\Delta=1.2$, $\lambdat=0.05$,
 $\theta(0)=0$, $d\theta/d\tau(0)=1$. 
Steady-state oscillations (see section \ref{sec:osci-ell}):
(c) $\epsilon=\pi/4$, $\Delta=0.1$, $\lambdat=0.1$,
 $\theta(0)=0$, $d\theta/d\tau(0)=1$; 
(d) $\epsilon=\pi/6$, $\Delta=0.001$, $\lambdat=0.1$,
 $\theta(0)=\pi/2$, $d\theta/d\tau(0)=-1.33\times 10^{-4}$.}  
  \label{fig:fig2}
\end{figure}

The perturbative results for elliptic motion of the suspension point
are compared to numerical solutions to (\ref{eq:eqell}) in figure
\ref{fig:fig2}.  Figure \ref{fig:fig2} (a) shows a contrarian
steady-state rotation, with $\epsilon=-1/2$ and $\Delta=0.1$.  Because
the oscillation amplitude is small we plotted $\theta(\tau)-\tau$
instead of $\theta(\tau)$.  The zeroth-order solution
(\ref{eq:zeroth-ell}) is then shown in the figure as
$\theta_0(\tau)-\tau = \Theta_0 \simeq -0.876$. As also seen there,
the first-order result (\ref{eq:first-ell}) is very close to the
numerical solution, and the second-order result (\ref{eq:second-ell})
agrees exactly with it.

For the contrarian motion with $\theta(0)=0$ shown in figure
\ref{fig:fig2} (a) we find steady rotations for $0.987 \leq
d\theta/d\tau(0) \leq 1.090$.  For other values of $d\theta/d\tau(0)$
the steady state is oscillatory, about $n\pi$ for some integer $n$.
Although not shown in the figure, we mention here that for
$\theta(0)=\pi/8$ and $\pi/4$ we could not find steady-state
rotations, any $d\theta/d\tau(0)$ leading to steady oscillations about
$n\pi$ instead.  If we set $\Delta$ and $\lambdat$ to the same values
as in figure \ref{fig:fig2} (a), but $\epsilon=+1/2$, then with
$\theta(0)=0$ steady-state rotations are reached for $0.559 \leq
d\theta/d\tau(0) \leq 1.519$, with $\theta(0)=\pi/8$ for $0.579 \leq
d\theta/d\tau(0) \leq 1.483$, and with $\theta(0)=\pi/4$ for $0.624
\leq d\theta/d\tau(0) \leq 1.419$.  These numerical results suggest
that the interval of initial angular velocities leading to
steady-state rotations decreases in length with increasing
$\theta(0)$.  Also, as intuitively expected, for fixed $\theta(0)$
that length is larger for direct than for contrarian motion.

In figure \ref{fig:fig2} (b) we show a direct steady-state rotation,
with $\epsilon=\pi/8$ and $\Delta=1.2$.  In this case the zeroth-order
result (\ref{eq:zeroth-ell}) is given by $\Theta_0 = -6\times
10^{-2}$.  Like in the case of linear excitation, we see that
perturbation theory converges for $\Delta\gtrsim 1$.  As seen in the
figure, the second-order result (\ref{eq:second-ell}) accurately
reproduces the numerical solution, and the third-order result
(\ref{eq:third-ell}) cannot be distinguished from the numerical one.
For these values of the parameters we found steady-state rotations for
any value of $\theta(0)$ and for any $0\leq d\theta/d\tau(0) \leq
10^3$. 

We show a steady-state oscillating solution in figure \ref{fig:fig2}
(c), with $\epsilon=\pi/4$, $\Delta=0.1=\lambdat$.  In this case
condition (\ref{eq:lambda-ell}) is violated, so steady rotations are
not possible.  The value of $\Theta_0 = 5\pi \simeq 15.708$ receives
corrections $\Delta \Theta_1$, $\Delta^3/6 \Theta_3$, so that the
solution oscillates about $\Theta \simeq 15.851$.  The accuracy of the
perturbative results from section \ref{sec:osci-ell} is apparent in
the figure, where no departure of the third-order result from the
numerical solution is visible.

In the case of linear excitation ($\epsilon=0$), as discussed in
section \ref{sec:numlin}, steady-state solutions oscillating about
$(2n+1)\pi/2$ could be found (see figure \ref{fig:fig1} (d)), although
finding them required some degree of fine-tuning of initial
conditions.  For $\epsilon\neq 0$ those solutions seem to be unstable.
An example is shown in figure \ref{fig:fig2} (d), with
$\epsilon=\pi/6$, where the solution initially oscillates about
$\Theta=\pi/2$ but moves away from that value to end up oscillating
about $\Theta=\pi$.  For smaller values of $\epsilon$ the growth of
the mean value of $\theta(\tau)$ from $\pi/2$ is slower, but even for
$\epsilon$ as low as $10^{-4}$ we could not find numerically a
stable solution of this type.

\section{Elliptically excited parametric pendulum}
\label{sec:gravell}

In the general case of elliptic motion of the suspension point, with
the $z$ axis chosen along the major axis of the ellipse and with
the direction of gravity at an angle $\pi+\alpha$ with that axis,
equation (\ref{eq:eqgral}) takes the form 
\begin{equation}
  \label{eq:eqgrav}
  \frac{d^2\theta}{d\tau^2} + \lambdat
  \frac{d\theta}{d\tau} + \frac{\Delta}{2}(1-\sin\epsilon)
  \sin(\theta+\tau) + \frac{\Delta}{2}
  (1+\sin\epsilon)\sin(\theta-\tau) + \Gamma \sin(\theta + \alpha) =
  0~.  
\end{equation}
This equation is invariant under the transformation $\theta\rightarrow
-\theta$, $\epsilon\rightarrow -\epsilon$, $\alpha\rightarrow
-\alpha$.  Thus, if $\theta(\tau,\epsilon,\alpha)$ is a solution to
(\ref{eq:eqgrav}) for some $\epsilon$, $\alpha$, then
$-\theta(\tau,-\epsilon,-\alpha)$ is also a solution for the same
$\epsilon$, $\alpha$.  There are no equilibrium solutions to
(\ref{eq:eqgrav}) when either $\epsilon\neq 0$ or $\alpha\neq0$.  For
$\epsilon=0=\alpha$ the equilibria are $\theta(\tau)=n\pi$, with
integer $n$, as in the case $\Gamma=0$.

When $\Delta$, $\lambdat$, $\Gamma\lesssim 1$ we can solve
(\ref{eq:eqgrav}) perturbatively for the steady-state solution along
the same lines as in the previous sections. The parameter $\Gamma$
defined in (\ref{eq:thinbar}) may be made arbitrarily small by tilting
the plane of the rotator with respect to the vertical, so that $g$
itself is small; or by making $\gamma$ large, $\gamma>\sqrt{g/L}$; or
by making small the rotator asymmetry ($(L_1-L_2)/L$ in the case of
the thin homogeneous bar discussed in section \ref{sec:rotat}).  For
arbitrary, but small, values of the parameters, we discuss below
steady-state rotations and two different oscillation regimes.
Furthermore, for $\Gamma\simeq 1/4$, there are two additional types of
solutions that we discuss in section \ref{sec:parres}.

\subsection{Steady-state rotations}
\label{sec:gravellrot}

The equation of motion (\ref{eq:eqgrav}) admits solutions of the form
(\ref{eq:solpert}), which we will denote by $\theta_{(g)}(\tau)$ to
distinguish them from the solutions $\theta(\tau)$ obtained in section
\ref{sec:ellipt} for $\Gamma=0$.  Like in that section, we need only
consider positive angular frequency solutions
$\theta_{(g)}(\tau,\epsilon,\alpha)$, $\epsilon>0$, describing direct
counterclockwise rotations.  Clockwise rotations are then given by
$-\theta_{(g)}(\tau,-\epsilon,-\alpha)$.  For the purpose of
perturbation theory we consider $\Delta$, $\lambdat$ and $\Gamma$ to
be small of the same order.  In particular, $\Gamma/\Delta$ and
$\lambdat/\Delta$ both remain finite as $\Delta\rightarrow 0$.  The
perturbative computation is a simple extension of the analyses of
sections \ref{sec:pert} and \ref{sec:ellrotat}, so we omit the details
for brevity, limiting ourselves to quoting the results through second
order.  At zeroth order we have
\begin{equation}
  \label{eq:zeroth-g}
  \theta_{(g)0}(\tau) = \theta_0(\tau)~,
\end{equation}
with $\theta_0(\tau)$ given by (\ref{eq:zeroth-ell}), so $\Theta_0$ is
independent of $\Gamma$ and relation (\ref{eq:lambda-ell}) holds
unchanged.  In particular, for circular excitation there can be no
contrarian motion, as in the case $\Gamma=0$.
At first order, 
\begin{equation}
  \label{eq:first-g}
  \theta_{(g)1}(\tau) = \theta_1(\tau) + \Gamma f_{(g)1}(\tau)~,  
\quad
  f_{(g)1}(\tau) = \sin(\tau + \Theta_0 + \alpha)~,
\end{equation}
with $\theta_1(\tau)$ given in (\ref{eq:first-ell}). At second
order we get, 
\begin{subequations}
  \label{eq:totsecond-g}
\begin{equation}
  \label{eq:second-g}
  \theta_{(g)2}(\tau) = \theta_{(g)1}(\tau) + \frac{1}{2} \Delta^2
  f_{2}(\tau) + \frac{1}{2} \Delta^2
  \Theta_{2} + \frac{1}{2}\Gamma^2 f_{(g)2}(\tau) + \frac{1}{2}
  \Gamma^2 \Theta_{(g)2}~,    
\end{equation}
with $f_2(\tau)$ and $\Theta_{2}$ as defined in (\ref{eq:second-ell}) and,
\begin{equation}
  \label{eq:pertgravalpha}
  \begin{aligned}
    \Theta_{(g)2} &= \left( \frac{3}{2} \tan(\Theta_0) - \frac{1}{8}
      \frac{1-\sin(\epsilon)}{1+\sin(\epsilon)} \frac{\sin(\Theta_0 +
        2\alpha)}{\cos(\Theta_0)}\right)~,
    \\
    f_{(g)2}(\tau) &= - \frac{3}{8} \frac{\Delta}{\Gamma}
    (1-\sin\epsilon) \sin(\tau-\alpha)
    + \frac{\Delta}{\Gamma} (1+\sin\epsilon) \sin(\tau+\alpha) \\
    &\quad+ \frac{5}{72} \frac{\Delta}{\Gamma} (1-\sin\epsilon)
    \sin(3\tau+2\Theta_0+\alpha)
    + \frac{1}{4} \sin(2\tau + 2 \Theta_0 + 2\alpha) ~.
  \end{aligned}
\end{equation}
\end{subequations}
Notice that in this equation we have eliminated $\lambdat$ in favour of
$\Theta_0$. 

A qualitative change with respect to the result of section
\ref{sec:ellipt} with $\Gamma=0$ is that the corrections $f_{(g)n}$
coming from gravity contain odd frequencies, so
$\theta_{(g)}(\tau)-\tau$ has period $2\pi$.  We notice also that the
phase of $\theta_{(g)}(\tau)$ acquires a dependence on $\Gamma$ at
second order.  The perturbative solution to (\ref{eq:eqgrav})
described by (\ref{eq:zeroth-g}) --- (\ref{eq:totsecond-g}) exists
only if (\ref{eq:lambda-ell}) is satisfied, due to the definition
(\ref{eq:zeroth-ell}) of $\Theta_0$.

\subsection{Steady-state oscillations ($\boldsymbol{\Gamma \sim \Delta}$)}
\label{sec:gravellosc}

We look for steady-state solutions of the form (\ref{eq:pertzero}).
As in the previous section, we consider $-\pi/2\leq\epsilon\leq\pi/2$
thus including the case of circular excitation.  For the purpose of
the perturbative calculation we assume $\Delta$, $\lambdat$,
$\Gamma\lesssim 1$ to be small of the same order.  From
(\ref{eq:eqgrav}) and (\ref{eq:pertzero}) we find the zeroth-order
solution
\begin{equation}
  \label{eq:gravosci-zero}
  \theta_{(g)0}(\tau) = \Theta_{(g)0} = n\pi - \alpha ~,
\end{equation}
with integer $n$.  The value of $\Theta_{(g)0}$ results from
consistency of the first-order equation, which then yields the
first-order solution
\begin{subequations}
  \label{eq:gravosci-first}
\begin{equation}
  \label{eq:gravosci-firsta}
  \theta_{(g)1}(\tau) = \Theta_{(g)0} + \Delta \Theta_{(g)1} + \Delta
  f_{(g)1}(\tau)~,  
\end{equation}
with $f_{(g)1}(\tau)$ given by (\ref{eq:first-osciell}) and, from the
requirement that $f_{(g)2}(\tau)$ be bounded, 
\begin{equation}
  \label{eq:gravosci-firstb}
  \Theta_{(g)1} = -\frac{1}{4} \left(\frac{\Delta}{\Gamma}\right)
  (\cos\epsilon)^2
  \frac{\sin(2\Theta_{(g)0})}{\cos(\Theta_{(g)0}+\alpha)}~.
\end{equation}
\end{subequations}
The second-order equation leads to
\begin{subequations}
  \label{eq:gravosci-scnd}
\begin{equation}
  \label{eq:gravosci-scnda}
 \theta_{(g)2}(\tau) = \theta_{(g)1}(\tau) + \frac{\Delta^2}{2}
 \Theta_{(g)2} + \frac{\Delta^2}{2} f_{(g)2}(\tau) ~,
\end{equation}
with
\begin{equation}
  \label{eq:gravosci-scndb}
  \begin{aligned}
\Theta_{(g)2} &= \frac{1}{\cos(\Theta_{(g)0}+\alpha)}
\left( \frac{\lambdat}{\Gamma} \sin\epsilon - \frac{1}{8}
  (\cos\epsilon)^2 \sin(\Theta_0 -\alpha) 
-\frac{3}{8} (\cos\epsilon)^2 \sin(3\Theta_{(g)0} + \alpha)  \right.\\
&\quad 
\left. 
- \frac{\Delta}{\Gamma}\Theta_{(g)1} (\cos\epsilon)^2
\cos(2\Theta_{(g)0}) 
\right)~,\\
    f_{(g)2}(\tau) &= \left(\Theta_{(g)1} -
      \lambdat/\Delta\right) (1+\sin\epsilon) \cos(\tau -
    \Theta_{(g)0}) + \left(\Theta_{(g)1} + \lambdat/\Delta\right)
    (1-\sin\epsilon) \cos(\tau + \Theta_{(g)0})\\ 
    &\quad + \frac{\Gamma}{2\Delta} \left( (1-\sin\epsilon)
      \sin(\tau-\alpha) - (1+\sin\epsilon) \sin(\tau+\alpha)
    \rule{0pt}{11pt}\right)\\ 
    &\quad + \frac{\Gamma}{2\Delta}\left( (1-\sin\epsilon)
    \sin(\tau+2\Theta_{(g)0}+\alpha) - 
    (1+\sin\epsilon) \sin(\tau-2\Theta_{(g)0}-\alpha)
  \rule{0pt}{11pt}\right)\\
    &\quad +\frac{1}{16}\left( (1-\sin\epsilon)^2
      \sin(2\tau+2\Theta_{(g)0})  -
      (1+\sin\epsilon)^2 \sin(2\tau-2\Theta_{(g)0})
    \rule{0pt}{11pt}\right)~,
  \end{aligned}
\end{equation}
\end{subequations}
where $\Theta_{(g)2}$ results from requiring consistency of the
third-order equation.

In the case $\epsilon=0=\alpha$, corresponding to linear excitation in
the same direction as gravity, $\theta_{(g)0}(\tau)$ is an exact
solution to (\ref{eq:eqgrav}), the equilibrium solution. In that case
the corrections $\Theta_{(g)1,2}$, $f_{(g)1,2}$ given in
(\ref{eq:gravosci-first}), (\ref{eq:gravosci-scnd}) vanish, as must
happen to all higher-order corrections $\Theta_{(g)i}$, $f_{(g)i}$
with $i\geq 2$.  For $\epsilon\neq0$ or $\alpha\neq0$ equations
(\ref{eq:gravosci-zero})--(\ref{eq:gravosci-scnd}) describe
steady-state oscillations with dimensionless fundamental angular
frequency 1 (in physical units, angular frequency $\gamma$) about a
central value $\Theta_{(g)} = \Theta_{(g)0} + \Delta \Theta_{(g)1} +
\Delta^2/2 \Theta_{(g)2} + \mathcal{O}(\Delta^3)$.  For $n$ even in
(\ref{eq:gravosci-zero}), the zeroth-order value $\Theta_0$
corresponds to the direction of gravity on the plane of the rotator.
By numerically solving (\ref{eq:eqgrav}) we find that the solutions
with $n$ odd are unstable in this regime $\Gamma\sim\Delta$, as
expected. 

This perturbative solution is generally valid for $\Delta$,
$\lambdat$, $\Gamma$ small, independently of whether condition
(\ref{eq:lambda-ell}) is satisfied.  If (\ref{eq:lambda-ell})
holds, the steady state rotations of section \ref{sec:gravellrot} also 
exist and the system may reach one or the other steady state
depending on its initial conditions.

\subsection{Steady-state oscillations ($\boldsymbol{\Gamma \ll \Delta}$)}
\label{sec:gravellosc2}

The perturbative solution
(\ref{eq:gravosci-zero})--(\ref{eq:gravosci-scnd}) given in the
previous section was obtained under the assumption
$\Gamma=\mathcal{O}(\Delta)$.  In particular, its limit as
$\Gamma\rightarrow 0$ is not well defined.  For $\Gamma \ll \Delta$
perturbation theory must be accordingly modified, which we do in this
section.  We focus our discussion on the specific case $\Delta<1$,
$\lambdat=\mathcal{O}(\Delta)$, $\Gamma=\mathcal{O}(\Delta^3)$.  (Many
other regimes are of course possible, which lead to results analogous
to those obtained here.)  With those assumptions for the parameters
and (\ref{eq:pertzero}) for $\theta_{(g)}(\tau)$, we solve
(\ref{eq:eqgrav}) perturbatively.  At zeroth order the solution is a
constant, $\theta_{(g)0}(\tau) = \Theta_{(g)0}$, to be determined.  At
order $\Delta$,
\begin{equation}
  \label{eq:gravosci-II-first}
  \theta_{(g)1}(\tau) = \Theta_{(g)0} + \Delta \Theta_{(g)1} + \Delta
  f_{(g)1}(\tau)~, 
\end{equation}
with $f_{(g)1}(\tau)$ given by (\ref{eq:first-osciell}), and
both $\Theta_{(g)0,1}$ still undetermined.  Consistency of the
$\mathcal{O}(\Delta^2)$ equation requires $\cos(\epsilon)^2
\sin(2\Theta_0) = 0$, so that for $\epsilon\neq\pm\pi/2$
\begin{equation}
  \label{eq:gravosci-II-zeroth}
  \Theta_{(g)0} = \frac{n}{2}\pi~,
\end{equation}
for some integer $n$.  With this, the second-order solution is 
\begin{equation}
  \label{eq:gravosci-II-secnd}
  \begin{gathered}
  \theta_{(g)2}(\tau) = \theta_{(g)1}(\tau) + \frac{\Delta^2}{2}
  (\Theta_{(g)2} + f_{(g)2}(\tau))~,\\
  \begin{aligned}
  f_{(g)2}(\tau) &= -\left(\frac{\lambdat}{\Delta} -
    \Theta_{(g)1}\right) (1+\sin\epsilon) \cos(\tau-\Theta_{(g)0})
  + \left(\frac{\lambdat}{\Delta} +
    \Theta_{(g)1}\right) (1-\sin\epsilon) \cos(\tau+\Theta_{(g)0})\\
  &\quad -\frac{1}{16} (1+\sin\epsilon)^2 \sin(2\tau-2\Theta_{(g)0})
    +\frac{1}{16} (1-\sin\epsilon)^2 \sin(2\tau+2\Theta_{(g)0})~.
  \end{aligned}
  \end{gathered}
\end{equation}
The third-order equation requires
\begin{equation}
  \label{eq:gravosci-II-theta1}
  \Theta_{(g)1} = \frac{1}{(\cos\epsilon)^2}
  \frac{1}{\cos(2\Theta_{(g)0})} \left( \frac{\lambdat}{\Delta}
    \sin\epsilon - 2 \frac{\Gamma}{\Delta^3} \sin(\Theta_{(g)0} +
    \alpha)\right)~,  
\end{equation}
for $f_{(g)3}$ to be bounded. The explicit expression for $f_{(g)3}$
is rather lengthy, so we omit it for brevity.
Finally, the $\mathcal{O}(\Delta^4)$ equation leads to the consistency
condition 
\begin{equation}
  \label{eq:gravosci-II-theta2}
  \Theta_{(g)2} =  -4 \frac{\Gamma}{\Delta^3}
  \frac{1}{(\cos\epsilon)^2} \frac{\cos(\Theta_{(g)0} +
    \alpha)}{\cos(2\Theta_{(g)0})} \Theta_{(g)1}~. 
\end{equation}
Equations (\ref{eq:gravosci-II-first})--(\ref{eq:gravosci-II-theta2})
yield a steady-state solution to (\ref{eq:eqgrav}) through second
order in $\Delta<1$.  Clearly, this perturbative solution breaks down
as $|\epsilon|$ approaches the upper bound $\epsilon_\mathrm{max}$
such that $\Theta_{(g)1}$ becomes $\sim1$.  Numerical study of these
oscillating solutions suggests that for $\epsilon_\mathrm{max}\lesssim
|\epsilon|\leq\pi/2$ they disappear from the spectrum of steady states.
Therefore, as intuitively expected, the oscillations described in this
section require an eccentric enough parametric excitation.

In the case of linear excitation perpendicular to gravity
($\epsilon=0$, $\alpha=\pm\pi/2$) the central value $\Theta_{(g)}$
about which the pendulum oscillates was obtained in
\cite{lan76,gra97,but01} for $\lambdat=0$.  Our result $\Theta_{(g)} =
\Theta_{(g)0} + \Delta \Theta_{(g)1} + \Delta^2/2 \Theta_{(g)2}$ given
by (\ref{eq:gravosci-II-zeroth}), (\ref{eq:gravosci-II-theta1}) and
(\ref{eq:gravosci-II-theta2}) agrees with those references at order
$\Delta$ if we set $\lambdat=0$ in (\ref{eq:gravosci-II-theta1}), but
further generalises it to the case $\lambdat>0$, to second order in
$\Delta$, to any angle $\alpha$, and to any large enough eccentricity
(i.e., small enough $\epsilon$).
The full solution
(\ref{eq:gravosci-II-first})--(\ref{eq:gravosci-II-theta2}), which
gives the steady-state oscillations about the central value
$\Theta_{(g)}$ through order $\Delta^2$, provides a similar
generalisation to the oscillations described in \cite{mic85} for an
inverted pendulum.  From (\ref{eq:gravosci-II-zeroth}) we see that our
results are applicable in the cases $\Theta_{(g)_0}=2k\pi$ with
integer $k$ (normal pendulum) and $\Theta_{(g)_0}=(2k+1)\pi$ (inverted
pendulum).  As discussed in the next section, by numerically solving
(\ref{eq:eqgrav}) the solutions with $\Theta_{(g)_0}=(2k+1)\pi/2$ are 
found to be unstable.

\subsection{Numerical results}
\label{sec:numgrav}

\begin{figure}[h]
  \centering
\begin{picture}(460,300)(0,0)
 \put(0,150){\scalebox{0.74}{\includegraphics{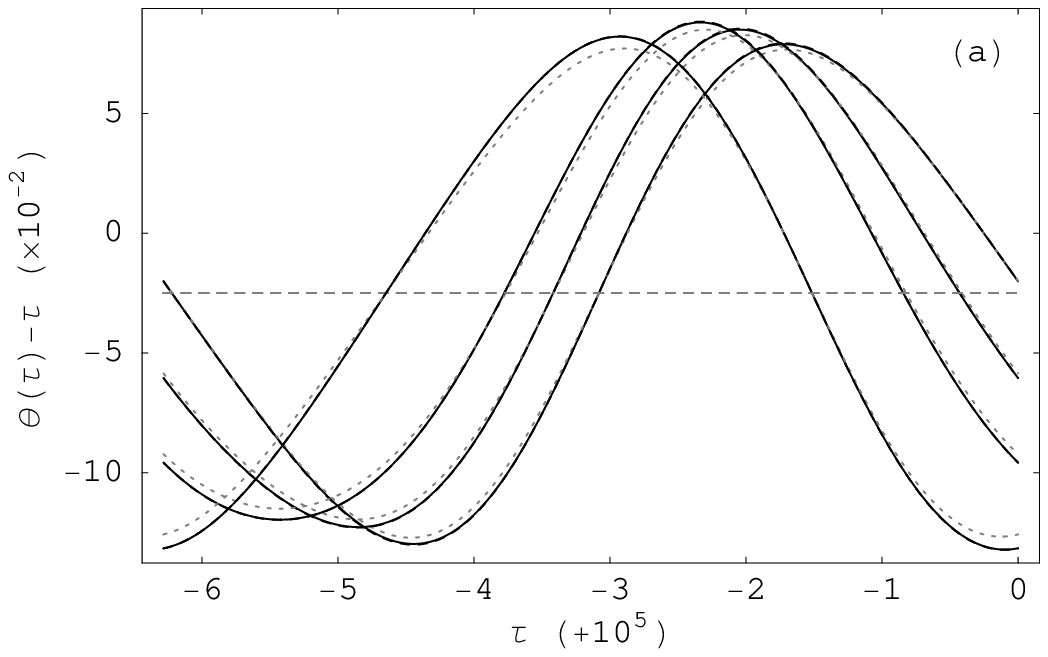}}}  
\put(80,300){\scalebox{0.74}{\rotatebox{-52}{\includegraphics{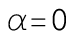}}}}
\put(80,300){\scalebox{0.74}{\rotatebox{-57}{\includegraphics{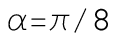}}}}
\put(84,270){\scalebox{0.74}{\rotatebox{-60}{\includegraphics{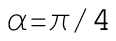}}}}
\put(72,260){\scalebox{0.74}{\rotatebox{-62}{\includegraphics{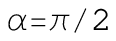}}}}
 \put(240,150){\scalebox{0.74}{\includegraphics{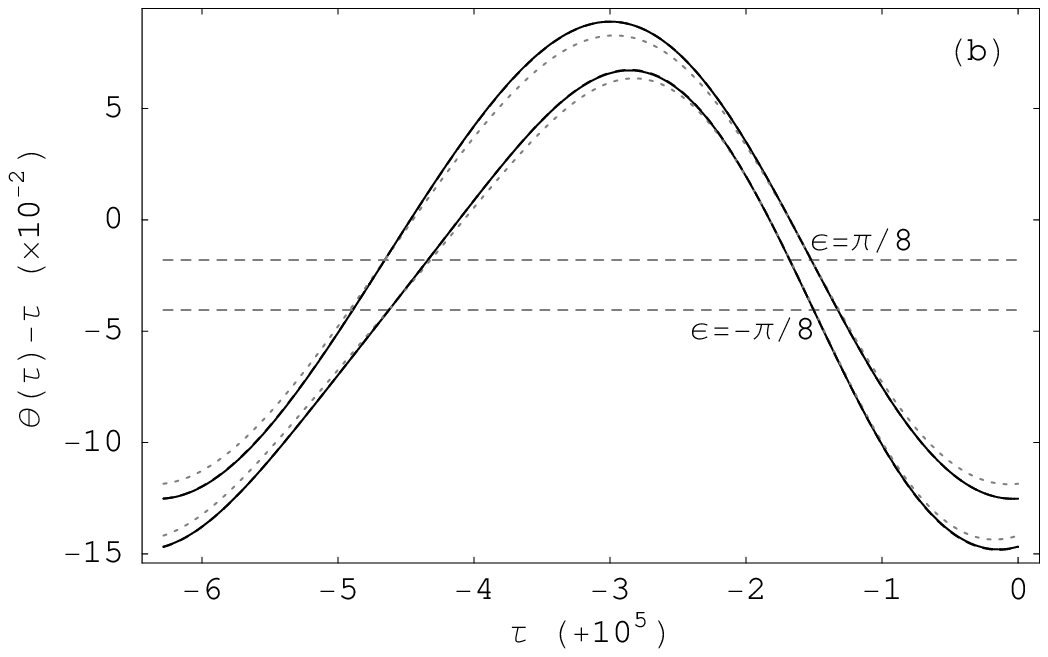}}}  
 \put(-7,0){\scalebox{0.76}{\includegraphics{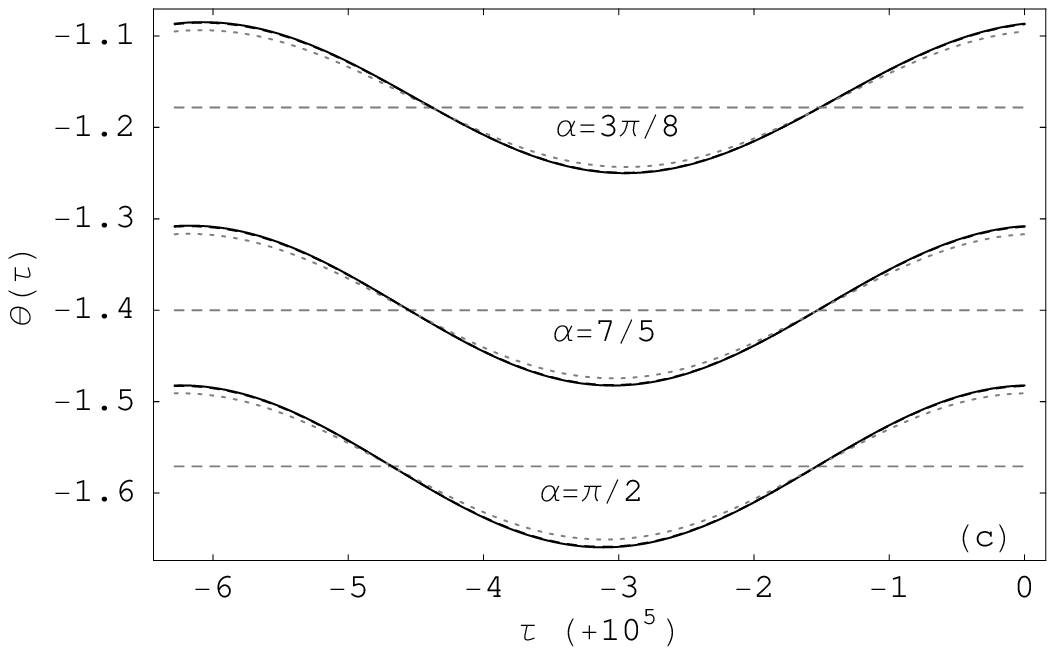}}}  
 \put(232,-0.1){\scalebox{0.765}{\includegraphics{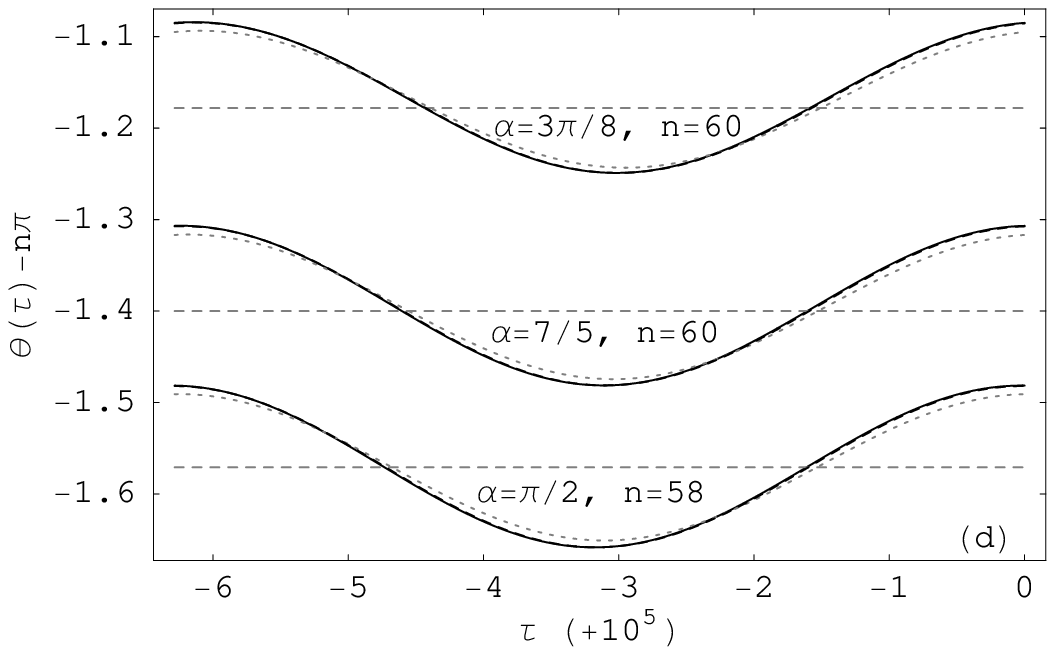}}}  
\end{picture}
  \caption{Numerical solution (black solid line) to equation
    (\ref{eq:eqgrav}), compared with perturbative solutions at zeroth
    (grey dashed line), first (grey dotted line) and second (black dashed
    line) orders as given in section
    \ref{sec:gravell}.  Steady-state rotations (see section
    \ref{sec:gravellrot}):
(a) $\epsilon=0$, $\Gamma=0.1$, $\Delta=0.08$,
$\lambdat=0.001$, $\theta(0)=0.1$, $d\theta/d\tau(0)=1.25$;
(b) $\epsilon=\pm\pi/8$, $\alpha=\pi/2$, $\Gamma=0.1$, $\Delta=0.08$,
$\lambdat=0.001$, $\theta(0)=0.1$, $d\theta/d\tau(0)=1.25$. 
Steady-state oscillations (see section \ref{sec:gravellosc}):
(c) $\epsilon=\pi/8$, $\Gamma=0.1$, $\Delta=0.08$,
$\lambdat=0.001$, $\theta(0)=0.1$, $d\theta/d\tau(0)=10^{-5}$;   
(d) $\epsilon=\pi/8$, $\Gamma=0.1$, $\Delta=0.08$,
$\lambdat=0.06$, $\theta(0)=0.1$, $d\theta/d\tau(0)=10$.}  
  \label{fig:fig3}
\end{figure}

The steady-state rotations obtained perturbatively in section
\ref{sec:gravellrot} are compared with numerical solutions to
(\ref{eq:eqgrav}) in figures \ref{fig:fig3} (a) and (b).  In figure
\ref{fig:fig3} (a) we show $\theta(\tau)-\tau$ for linear motion of
the suspension point ($\epsilon=0$), at several angles $\alpha$ with
the direction of gravity.  The zeroth-order result (\ref{eq:zeroth-g})
is $\Theta_{0} \simeq -0.025$, independently of $\Gamma$ and $\alpha$.
The first-order result (\ref{eq:first-g}) already accurately describes
the numerical solution, with which the second-order result
(\ref{eq:totsecond-g}) overlaps completely.  In figure \ref{fig:fig3}
(b) we show $\theta(\tau)-\tau$ for elliptic excitation with
$\epsilon=\pi/8$ perpendicular to gravity, both for direct and
contrarian motion.  The zeroth-order result (\ref{eq:zeroth-g}) is
$\Theta_{0} \simeq -0.018$ for direct and -0.040 for contrarian
motion.  The first and second-order results are seen to be in tight
agreement with the numerical solution.

In figures \ref{fig:fig3} (c) and (d) we show steady-state oscillations
as described in section \ref{sec:gravellosc}.  The parameters used in
figure \ref{fig:fig3} (c) do not violate the condition
(\ref{eq:lambda-ell}), but the initial angular velocity
$d\theta/d\tau(0)=10^{-5}$ is too small for steady-state rotations (at
$\epsilon=\pi/8$).  Numerically, for these parameters we find steady
rotations for $0.78\leq d\theta/d\tau(0)\leq 1.44$ for
$\alpha=3\pi/8$, $0.75\leq d\theta/d\tau(0)\leq 1.42$ for
$\alpha=7/5$, and $0.65\leq d\theta/d\tau(0)\leq 1.55$ for
$\alpha=\pi/2$.  The rotator oscillates about a central position
$\Theta_{(g)}$ whose zeroth-order value $\Theta_0=-\alpha$ is the
direction of gravity, but receives small second-order corrections
$\Delta^2/2 \Theta_{(g)2}=10^{-3}$, $6\times 10^{-4}$ and $10^{-5}$
for $\alpha=3\pi/8$, 7/5 and $\pi/2$, respectively. If instead of
$\epsilon=\pi/8$ we had chosen any other value
$-\pi/3\leq\epsilon\leq\pi/4$, the curves would vary by only a small
amount. For $|\epsilon|\gtrsim\pi/4$, however, steady rotations are
obtained even for the small initial angular velocity used here.
Notice that for $\epsilon=\pm\pi/2$ these oscillating solutions do not
exist, so we expect that as $\epsilon$ approaches those values the
oscillating steady states should disappear from the spectrum of
solutions. In figure \ref{fig:fig3} (d) we set the parameters to the
same values as in figure \ref{fig:fig3} (c) except for a larger
$\lambdat$, so that condition (\ref{eq:lambda-ell}) is violated, and a
larger initial angular velocity.  Except for the buildup of additional
turns during the transient, as indicated in the figure, the plots look
almost the same as those of figure \ref{fig:fig3} (c).  This is due to
the fact that $\lambdat$ enters the solutions of section
\ref{sec:gravellosc} only at order $\Delta^2$.

\begin{figure}[h]
  \centering
\begin{picture}(460,300)(0,0)
 \put(0,150){\scalebox{0.74}{\includegraphics{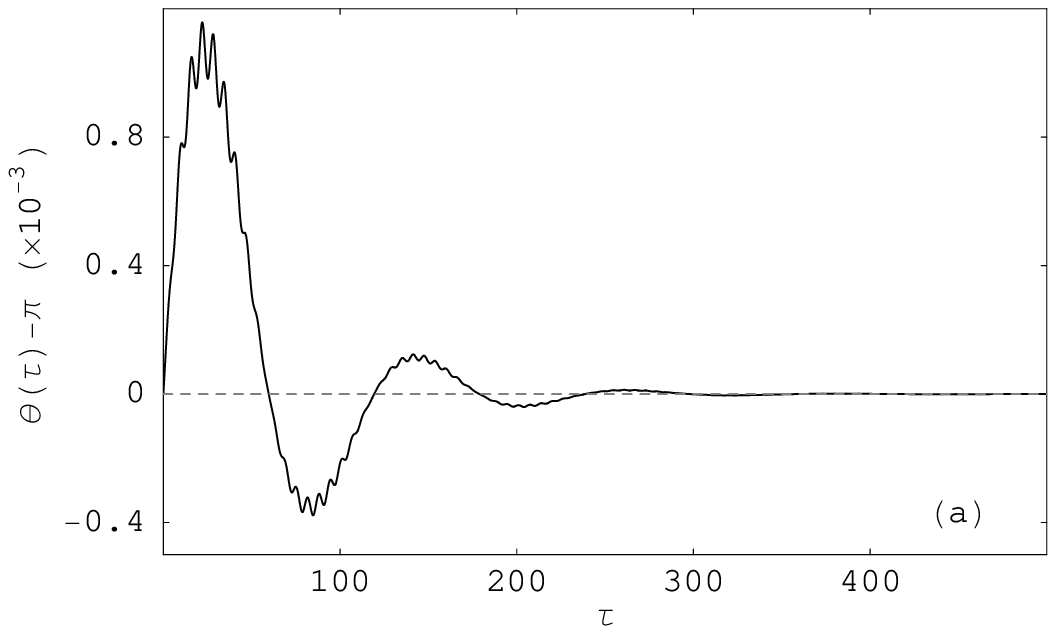}}}  
 \put(240,150){\scalebox{0.73}{\includegraphics{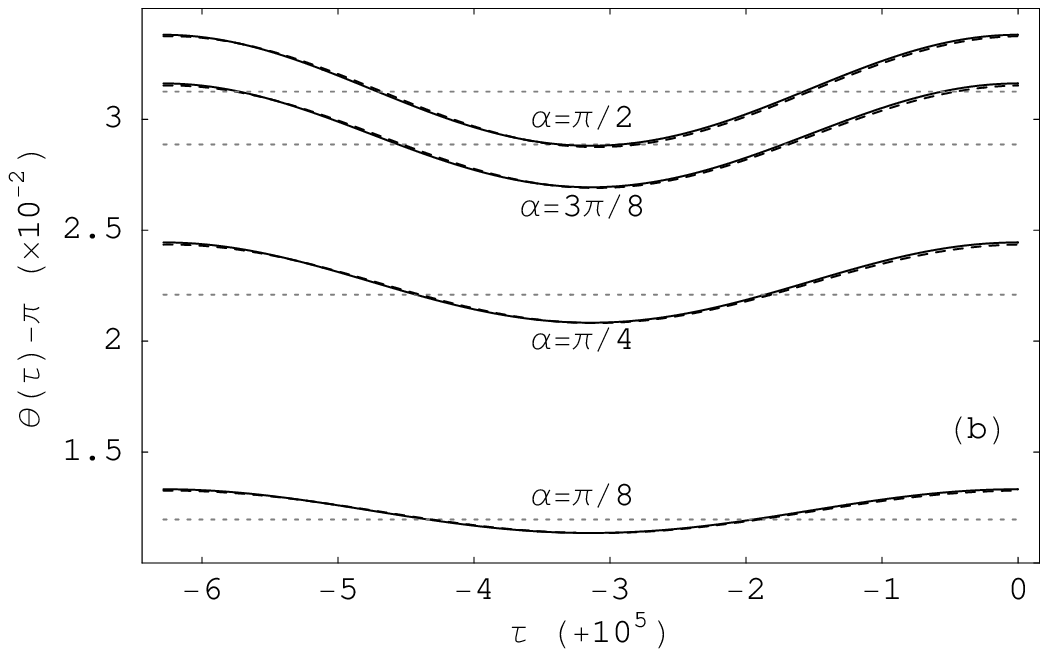}}}  
 \put(7,0){\scalebox{0.725}{\includegraphics{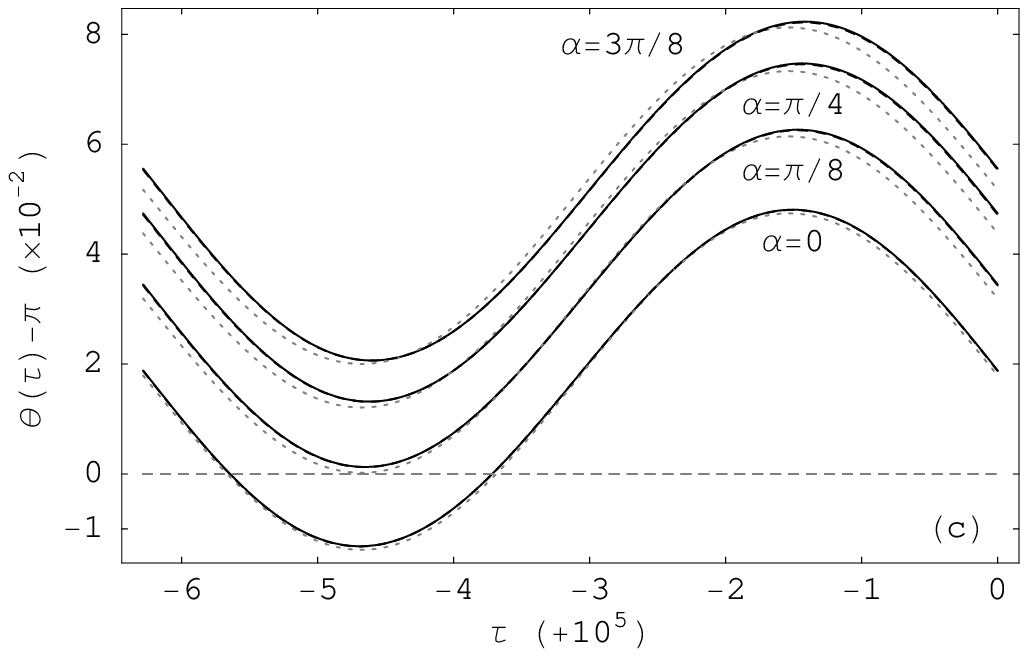}}}  
 \put(245.5,0){\scalebox{0.725}{\includegraphics{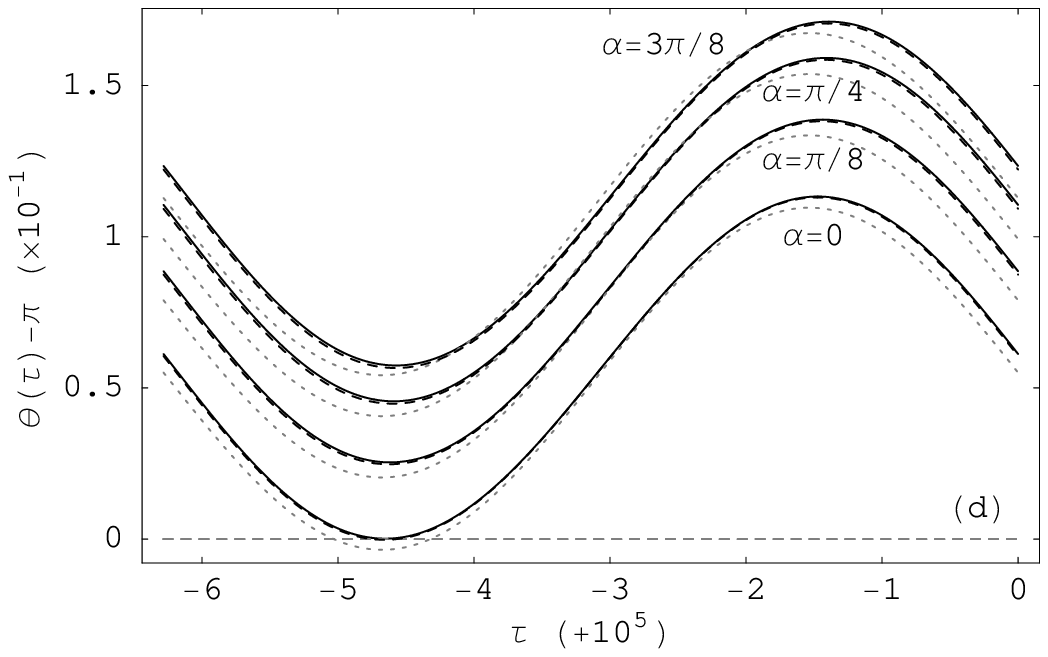}}}  
\end{picture}
  \caption{Inverted-pendulum solutions to equation
    (\ref{eq:eqgrav}). Numerical solution (black solid line),
    compared with perturbative solutions at zeroth 
    (grey dashed line), first (grey dotted line) and second (black dashed
    line) orders as given in section  \ref{sec:gravellosc2}.  The
    parameters are set to $\Gamma=10^{-4}$, $\Delta=0.08$,
    $\lambdat=0.0375$, $\theta(0)=\pi$, $d\theta/d\tau(0)=10^{-4}$, and
    (a) $\epsilon=0$, $\alpha=0$; 
    (b) $\epsilon=0$;
    (c) $\epsilon=\pi/8$;
    (d) $\epsilon=\pi/4$.}
  \label{fig:fig4}
\end{figure}

The oscillating solutions for $\Gamma\ll \Delta$ obtained in section
\ref{sec:gravellosc2} are illustrated in figure \ref{fig:fig4}.  We
focus there on solutions with $\Theta_{(g)0} = (2n+1)\pi$, describing
the well-known phenomenon of the ``inverted pendulum'' (see the
references cited in section \ref{sec:intro} above). In the case
of linear parametric excitation parallel to gravity
($\epsilon=0=\alpha$), the exact solutions to (\ref{eq:eqgrav})
$\theta(\tau)=n\pi$ with $n$ odd become stable for $\Gamma\ll \Delta$.
In figure \ref{fig:fig4} (a) we show one such solution, including the
transient and the relaxation to the steady state $\theta(\tau)=\pi$.
When $\epsilon=0\neq\alpha$ there are no equilibria.  In figure
\ref{fig:fig4} (b) we show the oscillating steady state in that case,
for various values of $\alpha$.  For $\epsilon=0$ and
$\Theta_{(g)0}=\pi$ the first-order result
(\ref{eq:gravosci-II-first}) becomes constant, as shown in the figure,
whereas the second-order result (\ref{eq:gravosci-II-secnd})
accurately reproduces the numerical solution.  The pendulum oscillates
about $\Theta_{(g)} = \Theta_{(g)0} + \Delta \Theta_{(g)1} +
\Delta^2/2 \Theta_{(g)2}+\ldots$$\simeq \pi+1.2\times10^{-2}$,
$\pi+2.3\times 10^{-2}$, $\pi+2.9\times 10^{-2}$, $\pi+3.1\times
10^{-2}$, for $\alpha=\pi/8$, $\pi/4$, $3\pi/8$, $\pi/2$,
respectively.  We recall here that $\Theta$ is measured with respect
to the trajectory of the suspension point, not that of gravity.  For
example, when $\alpha=\pi/2$, figure \ref{fig:fig4} (b) shows that the
pendulum is oscillating about an almost horizontal position.  Similar
results are obtained in the case of elliptic excitation, as shown in
figures \ref{fig:fig4} (c) and (d) with $\epsilon=\pi/8$ and
$\epsilon=\pi/4$.  In those figures we omitted the curves corresponding
to $\alpha=\pi/2$ for clarity, because they lie very close to those
for $\alpha=3\pi/8$.  As can be seen in the figure, the oscillation
amplitude depends on both $\alpha$ and $\epsilon$.  For
$\epsilon\gtrsim\pi/4$ higher orders beyond the second are needed to
reproduce the numerical solutions and, as $\epsilon$ approaches
$\pm\pi/2$, perturbation theory becomes inapplicable.  For
$\epsilon=\pm\pi/2$ the oscillating steady states described in sections
\ref{sec:gravellosc2} do not exist.

Finally, we notice that (\ref{eq:gravosci-II-zeroth}) allows solutions
oscillating about $\Theta_{(g)0} = (2n+1)\pi/2$.  Numerical study
shows that, for the values of $\Gamma$, $\Delta$, $\lambdat$
considered in figure \ref{fig:fig4}, those solutions are unstable even
for $\epsilon=0$, and therefore they cannot describe steady states.
We have also numerically solved (\ref{eq:eqgrav}) for many sets of
parameter values in the regime $\Gamma\ll\Delta\ll\lambdat$, with
$\epsilon=0$ and several angles $\alpha$, and in all cases found the
solutions to be unstable.  That instability is of the same type as in
the case $\Gamma=0$, as shown in figure \ref{fig:fig2} (d): the central
value $\Theta_{(g)}$ about which $\theta_{(g)}(\tau)$ oscillates is
not constant, but its $\tau$ dependence is slow compared to the period
of oscillation.  Thus, the oscillations with $\Theta_{(g)0} =
(2n+1)\pi/2$ are well described by the perturbative solutions of
section \ref{sec:gravellosc2}, but only up to an additive constant and
for time intervals short compared with the time scales of variation of
the central value $\Theta_{(g)}$.

\section{Parametric resonance}
\label{sec:parres}

Besides the steady-state solutions obtained in section
\ref{sec:gravell}, when the value of $\Gamma$ is in the vicinity of
$(\ell/2)^2$, $\ell$ integer, (\ref{eq:eqgrav}) possesses resonant
solutions.  In this section we describe in detail the first parametric
resonance mode, with $\Gamma \simeq 1/4$.
We assume $\Delta\lesssim 1$, $\lambdat=\mathcal{O}(\Delta)$ and
$\Gamma=1/4 + \delta$ with $\delta=\mathcal{O}(\Delta)$, and expand in
powers of $\Delta^{1/2}$ to find a solution of the form
\begin{equation}
  \label{eq:parressol}
  \theta(\tau) = \Theta_0 + \sum_{k\geq 1} \Delta^{k/2} f_k(\tau)~. 
\end{equation}
In this case it turns out to be more convenient not to separate
explicitly from $f_k(\tau)$ an additive constant term, as done in the
expansions (\ref{eq:solpert}), (\ref{eq:pertzero}). 
Substituting (\ref{eq:parressol}) into (\ref{eq:eqgrav}), at
zeroth order we find $\sin(\Theta_0+\alpha)=0$.  The zeroth-order
solution is then,
\begin{equation}
  \label{eq:parres-zeroth}
  \theta_0(\tau) = \Theta_0 = -\alpha + n\pi~,
\end{equation}
with integer $n$.  At order $\Delta^{1/2}$ the equation of motion
(\ref{eq:eqgrav}) yields
\begin{equation}
  \label{eq:parres-first-eq}
  \frac{d^2f_1}{d\tau^2} + \frac{1}{4} \cos(n\pi) f_1(\tau) = 0~.
\end{equation}
Requiring $f_1$ to be bounded leads to the condition that $n$ in
(\ref{eq:parres-zeroth}) must be even, so the pendulum oscillates at
resonance about its position of lowest potential energy.  The
first-order solution is then,
\begin{equation}
  \label{eq:parres-first-sol}
  \begin{gathered}
    \theta_1(\tau) = \Theta_0 + \Delta^{1/2} f_1(\tau)~,
\qquad
    f_1(\tau) = \rho_1 \cos(\tau/2 - \varphi_1)~,\\
  \end{gathered}
\end{equation}
with the constants $\rho_1$, $\varphi_1$ undetermined at this order.
Similarly, the equation of motion at order $\Delta$ is
\begin{equation}
  \label{eq:parres-scnd-eq}
  \frac{d^2f_2}{d\tau^2} + \frac{1}{4} f_2(\tau) = 
  -\frac{1}{2} (1-\sin\epsilon) \sin(\tau-\alpha)  
  +\frac{1}{2} (1+\sin\epsilon) \sin(\tau+\alpha)~,  
\end{equation}
leading to the second-order solution
\begin{equation}
  \label{eq:parres-scnd-sol}
  \begin{gathered}
    \theta_2(\tau) = \theta_1(\tau) + \Delta f_2(\tau)~, \\   
    f_2(\tau) = \rho_2 \cos(\tau/2 - \varphi_2) + \frac{2}{3}
    (1-\sin\epsilon) \sin(\tau-\alpha) - \frac{2}{3}
    (1+\sin\epsilon) \sin(\tau+\alpha)~,
  \end{gathered}
\end{equation}
with the constants $\rho_2$, $\varphi_2$ undetermined at this order.
At $\mathcal{O}(\Delta^{3/2})$ we obtain the equation for $f_3(\tau)$
\begin{equation}
  \label{eq:parres-thrd-eq}
  \begin{aligned}
  \frac{d^2f_3}{d\tau^2} + \frac{1}{4} f_3(\tau) &=
  -\frac{\lambdat}{\Delta} \frac{df_1}{d\tau} - \frac{\delta}{\Delta}
  f_1(\tau) + \frac{1}{24} f_1(\tau)^3 - \frac{1}{2} (1-\sin\epsilon)
  \cos(\tau-\alpha) f_1(\tau)\\
  &\quad- \frac{1}{2} (1+\sin\epsilon)
  \cos(\tau+\alpha) f_1(\tau)~.
  \end{aligned}
\end{equation}
The right-hand side of this equation, with $f_1(\tau)$ given by
(\ref{eq:parres-first-sol}), contains terms proportional to
$\cos(\tau/2)$ and $\sin(\tau/2)$ leading to an unbounded solution
$f_3(\tau)$.  Requiring those terms to vanish determines the constants
$\rho_1$, $\varphi_1$ in (\ref{eq:parres-first-sol}) to be
\begin{equation}
  \label{eq:rho1-phi1}
  \begin{aligned}
  \rho_1 &= 4 \sqrt{2 \delta/\Delta + \sqrt{(\cos\alpha)^2
      + (\sin\epsilon)^2 (\sin\alpha)^2
      - (\lambdat/\Delta)^2 }}~,
\\
\varphi_1 &= \arctan\left(
  \frac{-\cos\alpha + \sqrt{(\cos\alpha)^2 +
  (\sin\epsilon)^2(\sin\alpha)^2
  - (\lambdat/\Delta)^2}}{\lambdat/\Delta-\sin\epsilon \sin\alpha} \right)
  + n_{\varphi_1} \pi~,
  \end{aligned}
\end{equation}
where $n_{\varphi_1}$ is an integer parameter on which the solution
depends, similarly to the parameter $n$ in (\ref{eq:parres-zeroth}).
We remark, however, that from (\ref{eq:parres-first-sol}) it is
apparent that the solutions depend only on whether $n_{\varphi_1}$ is
even or odd.  With (\ref{eq:rho1-phi1}), equation
(\ref{eq:parres-thrd-eq}) leads to a third-order solution of the form, 
\begin{equation}
  \label{eq:parres-thrd-sol}
  \begin{aligned}
    \theta_3(\tau) &= \theta_2(\tau) + \Delta^{3/2} f_3(\tau)~,\\
  f_3(\tau) &= A_3 \sin(\tau/2) + B_3 \cos(\tau/2) 
-\frac{1}{192} \rho^3 \cos(3\tau/2-3 \varphi_1)  +\frac{1}{8}  \rho
(1-\sin\epsilon)\cos(3\tau/2-\alpha -\varphi_1) \\
&\quad
+\frac{1}{8} \rho (1+\sin\epsilon) \cos(3\tau/2+\alpha -\varphi_1)~.  
  \end{aligned}
\end{equation}
The coefficients $A_3$, $B_3$ in $f_3(\tau)$ are determined by the
requirement that the perturbative equation of order $\Delta^{5/2}$
must have a bounded solution.  The resulting expressions are too
lengthy to transcribe here, so we omit them (see, however, Appendix
\ref{sec:appa}).  Finally, consistency of the equation at order
$\Delta^2$, which determines $f_4(\tau)$, requires
\begin{equation}
  \label{eq:rho2}
  \rho_2=0~,
\end{equation}
which fixes the second-order solution $\theta_2(\tau)$ in
(\ref{eq:parres-scnd-sol}). Summarising, the steady-state solution at
order $\Delta^0$ is given by (\ref{eq:parres-zeroth}); at order
$\Delta^{1/2}$ by (\ref{eq:parres-first-sol}) with
(\ref{eq:rho1-phi1}); at order $\Delta^1$ by
(\ref{eq:parres-scnd-sol}) with (\ref{eq:rho2}); and at order
$\Delta^{3/2}$ by (\ref{eq:parres-thrd-sol}).

We remark that $f_1(\tau)$ oscillates with angular frequency 1/2, and
$f_2(\tau)$ with angular frequency 1.  $f_{1,2}(\tau)$ determine the
equations for the higher-order corrections $f_k(\tau)$, $k>2$.  The
structure of those perturbative equations is such that $f_k(\tau)$
contains only terms with half-integer (respectively integer) angular
frequency if $k$ is odd (respectively even).  As a consequence, the
following relations, which are easily checked for the low-order
solutions given above, actually persist to all orders of perturbation
theory 
\begin{equation}
  \label{eq:symmeq}
  \theta(\tau,\epsilon,\alpha,n_{\varphi_1}+1) = 
  -\theta(\tau,-\epsilon,-\alpha,n_{\varphi_1}) + 2n\pi = 
  \theta(\tau+2\pi,\epsilon,\alpha,n_{\varphi_1})~. 
\end{equation}
From the second equality we immediately see that in the case of linear
motion of the suspension point parallel to gravity,
$\epsilon=0=\alpha$, all terms with integer angular frequencies must
vanish.  In that case only the terms with odd $k$ contribute to
(\ref{eq:parressol}).  In general, for arbitrary values of $\epsilon$
and $\alpha$, equations (\ref{eq:parres-zeroth})--(\ref{eq:rho2}) describe
two different solutions according to whether $n_{\varphi_1}$ is even
or odd.  Those two solutions are related, modulo 2$\pi$, by the
symmetries of (\ref{eq:eqgrav}) as shown by (\ref{eq:symmeq}).  Notice
that, whereas the invariance of (\ref{eq:eqgrav}) under $\theta(\tau)
\rightarrow \theta(\tau + 2\pi)$ did not play any role in section
\ref{sec:gravell} because there only integer angular frequencies
appear, that symmetry acts non-trivially on the resonant solutions
described in this section.

We see from (\ref{eq:rho1-phi1}) that there are two necessary
conditions for these perturbative solutions to exist.  The first one
places an upper bound on damping for a given excitation amplitude,
\begin{equation}
  \label{eq:firstcond}
  \lambdat \leq \Delta \sqrt{(\cos\alpha)^2 + (\sin\epsilon)^2
  (\sin\alpha)^2}~. 
\end{equation}
Thus, for linear excitation ($\epsilon=0$) there is no parametric
resonance if $\alpha=\pm\pi/2$.  For circular excitation the
right-hand side
of (\ref{eq:firstcond}) becomes independent of $\alpha$, as
expected. The second condition,
\begin{equation}
  \label{eq:secncond}
  \delta > -\frac{1}{2}\Delta \sqrt{(\cos\alpha)^2 + (\sin\epsilon)^2
  (\sin\alpha)^2 - (\lambdat/\Delta)^2}~,
\end{equation}
puts a (negative) lower bound on $\delta = \Gamma-1/4$.  Notice that
there is, apparently, no restriction on how large $\delta$ can be as
long as it is positive.  This is, in fact, an artifact of perturbation
theory: a non-perturbative stability analysis leads to well defined
limits on the magnitude of $\delta$ for resonance to be possible
\cite{arn78}.  Yet, perturbation theory does restrict how negative
$\delta$ can be through (\ref{eq:secncond}).  Numerical study confirms
this asymmetric situation: given $0<\delta_0\lesssim\Delta$ such that
$-\delta_0$ violates (\ref{eq:secncond}), resonant solutions can be
found numerically for $\delta=\delta_0$, but not for
$\delta=-\delta_0$.

\subsection{Steady-state rotations with angular velocity 1/2}
\label{sec:angfreqhalf}

When $\Gamma=1/4+\delta$, with $\delta=\mathcal{O}(\Delta)$,
$\lambdat\lesssim \Delta$ and $\Delta\ll 1$, equation
(\ref{eq:eqgrav}) can be seen as a small perturbation about the
unperturbed equation
\begin{equation}
  \label{eq:unpert}
  \frac{d^2\theta}{d\tau^2} + \omega^2 \sin(\theta(\tau))=0~,
\qquad
  \omega=1/2~.
\end{equation}
In that case there must be solutions to (\ref{eq:eqgrav}) close to
those of (\ref{eq:unpert}), with corrections of $\mathcal{O}(\Delta)$.
For $\omega^2=1/4$, (\ref{eq:unpert}) possesses solutions of the form
$\theta(\tau) = \tau/2 + \Theta_0 + \sin(\tau/2+\Theta_0) + \ldots$,
for some constant $\Theta_0$, that can be computed perturbatively as
an expansion in powers of $\omega^2$.  How the remaining terms in
(\ref{eq:eqgrav}) are added to (\ref{eq:unpert}) as perturbations
depends on their relative size as compared to $\omega^2=1/4$.  For
$\Delta\sim 10^{-2}$--$10^{-1}$, $\lambdat\sim 10^{-3}$--$10^{-2}$,
$\delta\sim \Delta$, an appropriate expansion is obtained by setting
$\Delta=\mathcal{O}(\omega^4)$, $\delta=\mathcal{O}(\omega^4)$ and
$\lambda=\mathcal{O}(\omega^6)$. Thus, we expand (\ref{eq:eqgrav})
perturbatively, in a completely analogous way as in the previous
sections, with
\begin{equation}
  \label{eq:slophlfpert}
  \theta(\tau) = \frac{1}{2}\tau+\Theta_0 + \sum_{n=1}^\infty
  (\omega^2)^n (\Theta_n+f_n(\tau))~.
\end{equation}
At zeroth order we obtain, 
\begin{equation}
  \label{eq:slophlfzeroth}
  \theta_0(\tau) = \frac{1}{2}\tau+\Theta_0~,
\qquad
  \Theta_0 = -\frac{1}{2} \arcsin\left(\frac{\lambdat}{\Delta(1+\sin\epsilon)}\right) 
    -\frac{1}{2} \alpha + n\pi~,
\end{equation}
with $n$ integer.  We see that these steady states exist only if
\begin{equation}
  \label{eq:slophlflambda}
  \lambdat \leq \Delta (1+\sin\epsilon)~.
\end{equation}
At first order we get,
\begin{equation}
  \label{eq:slophlffirst}
  \theta_1(\tau) = \theta_0(\tau) + \frac{1}{4} (\Theta_1 +
  f_1(\tau))~,
\quad
  \frac{1}{4}\Theta_1 = -2 \delta \tan(2\Theta_0+\alpha)~,
\quad
  \frac{1}{4} f_1(\tau) = \sin(\tau/2+\Theta_0+\alpha)~.
\end{equation}
We remark that this first-order solution already depends on all of the
parameters in (\ref{eq:eqgrav}), and therefore on all of the physical
quantities entering it: elliptic parametric excitation and gravity,
their relative angle, and damping.  Notice also that at this order
(\ref{eq:slophlffirst}) is a solution to (\ref{eq:unpert}), the
additional terms in (\ref{eq:eqgrav}) entering (\ref{eq:slophlffirst})
only through $\Theta_{0,1}$.  At second order the perturbative
solution is
\begin{equation}
  \label{eq:slophlfsecond}
  \begin{gathered}
  \theta_2(\tau) = \theta_1(\tau) + \frac{1}{16} (\Theta_2 +
  f_2(\tau))~,  \\
\frac{1}{16} \Theta_2 =
-\frac{1}{\cos(2\Theta_0+\alpha)}\frac{1}{1+\sin\epsilon}
\frac{\lambdat}{4\Delta} - \delta \Theta_1 + \frac{1}{4}
\tan(2\Theta_0 + \alpha) + \frac{1}{16} \Theta_1^2
\tan(2\Theta_0+\alpha)~, \\ 
\begin{aligned}
  \frac{1}{16} f_2(\tau) &= \frac{1}{8} \sin(\tau+2\Theta_0+2\alpha)
  -2\Delta (1+\sin\epsilon)
\sin(\tau/2-\Theta_0) + \frac{2}{9} \Delta (1-\sin\epsilon) \sin(3/2
\tau + \Theta_0) \\
 &\quad + 4\delta \sin(\tau/2 + \Theta_0 + \alpha) + \frac{1}{4}
 \Theta_1 \cos(\tau/2+\Theta_0+\alpha)~.
\end{aligned}
  \end{gathered}
\end{equation}
Because the expansion parameter $\omega^2=1/4$ is not very small, the
third-order correction is necessary in order to obtain a perturbative
solution as accurate as those given in previous sections.  We take
into account the third-order terms in numerical computations in the
following section, but we omit them here for brevity.

\subsection{Numerical results}
\label{sec:parresnumres}

\begin{figure}
  \centering
\begin{picture}(460,450)(0,0)
 \put(-2.5,299.5){\scalebox{0.748}{\includegraphics{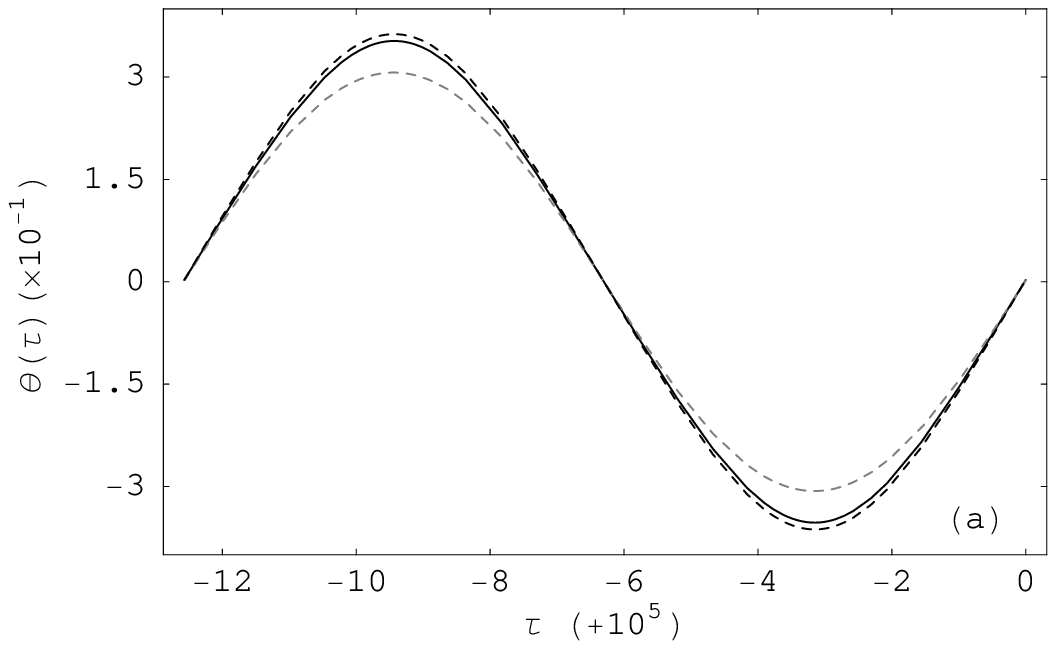}}}  
 \put(240,300){\scalebox{0.74}{\includegraphics{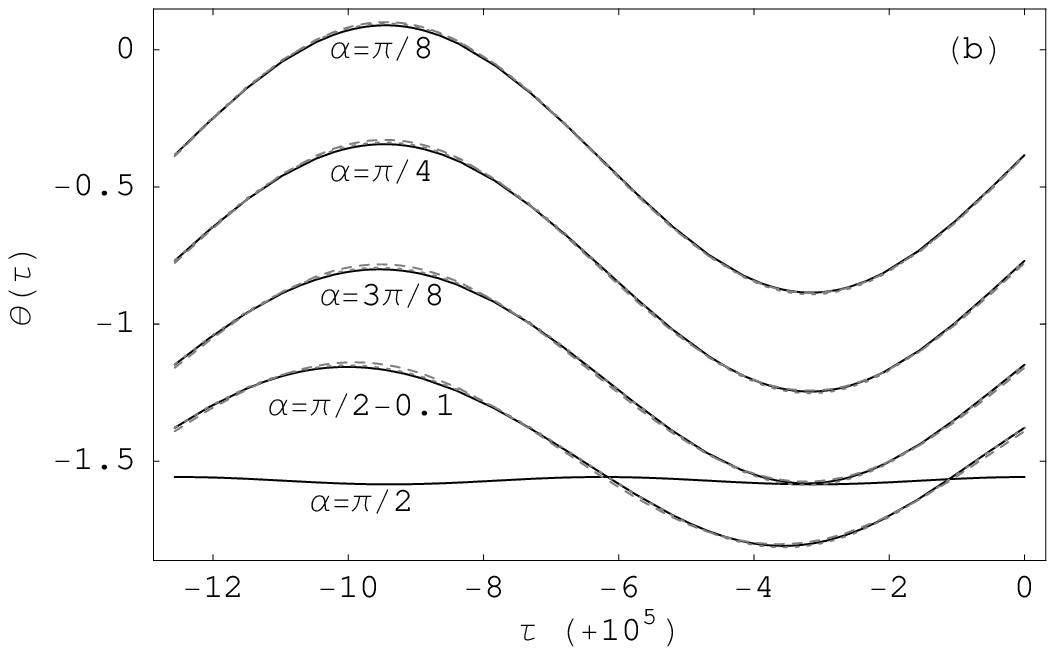}}}  
 \put(0,150){\scalebox{0.74}{\includegraphics{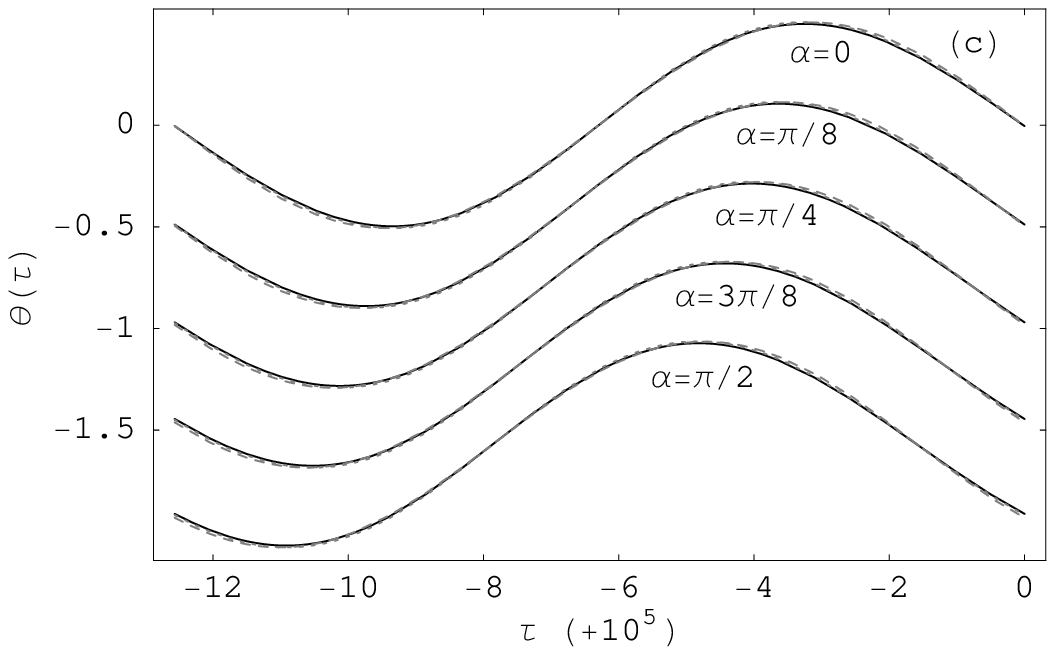}}}  
 \put(244.5,151){\scalebox{0.735}{\includegraphics{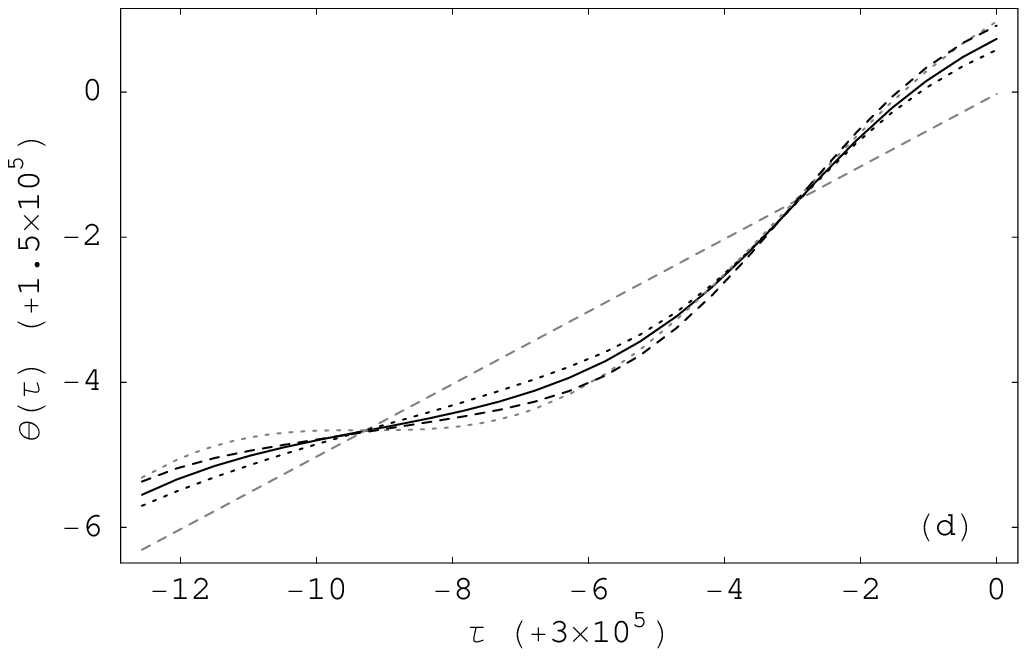}}}  
 \put(4,0){\scalebox{0.735}{\includegraphics{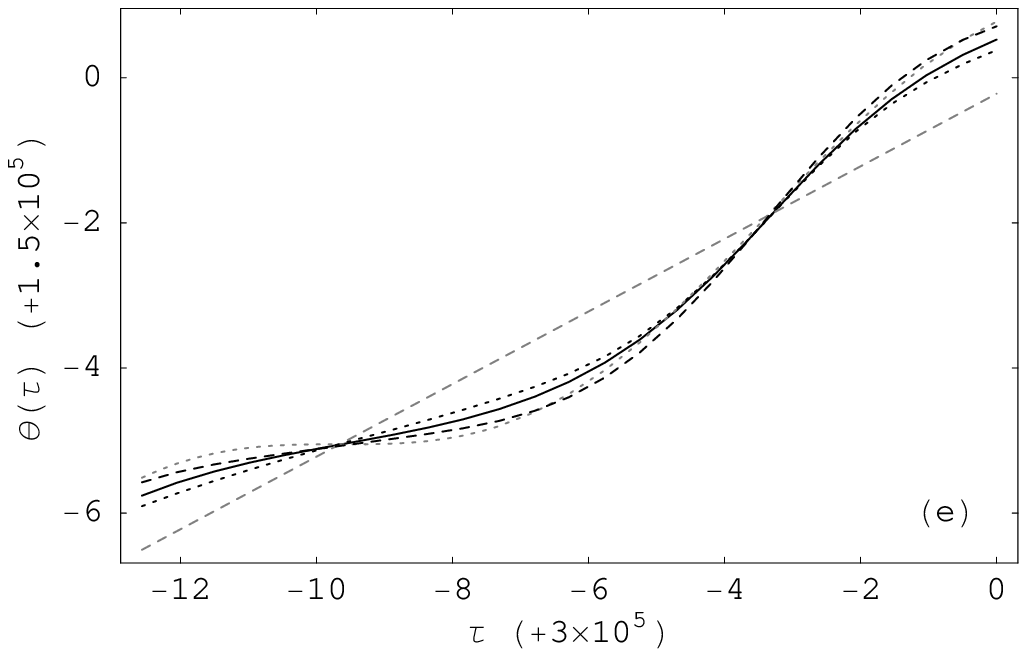}}}  
 \put(244.5,0){\scalebox{0.735}{\includegraphics{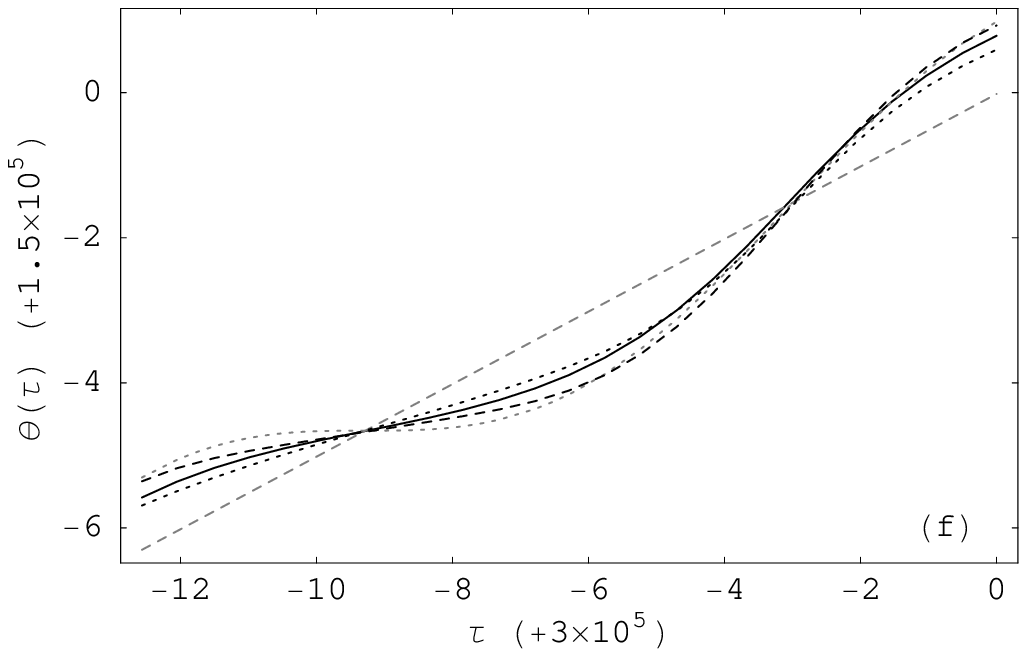}}}  
\end{picture}
  \caption{Solutions to equation (\ref{eq:eqgrav}) in the
    parametric-resonance region. Numerical solution (black solid line),
    compared with perturbative solutions at first
    (grey dashed line), second (grey dotted line) and third (black dashed
    line) orders as given in section  \ref{sec:parres}.  The
    parameters are set to 
    (a) $\epsilon=0$, $\alpha=0$, $n_{\varphi_1}=0$, $\Gamma=0.203$, $\Delta=0.1$,
    $\lambdat=5\times10^{-3}$, $\theta(0)=0$,
    $d\theta/d\tau(0)=-10^{-4}$;
    (b) $\epsilon=0$, $n_{\varphi_1}=0$, $\Gamma=0.253$, $\Delta=0.01$,
    $\lambdat=0.005$, $\theta(0)=0$, $d\theta/d\tau(0)=0.01$;
    (c) $\epsilon=\pi/2$, $n_{\varphi_1}=1$, $\Gamma=0.253$, $\Delta=0.01$,
    $\lambdat=0.005$, $\theta(0)=0$, $d\theta/d\tau(0)=0.1$; 
    (d) $\epsilon=0$, $\alpha=0$, $\Gamma=0.243$, $\Delta=0.01$,
    $\lambdat=5\times10^{-4}$, $\theta(0)=0$, $d\theta/d\tau(0)=1$;
    (e) $\epsilon=0$, $\alpha=\pi/8$, $\Gamma=0.253$, $\Delta=0.01$,
    $\lambdat=5\times10^{-4}$, $\theta(0)=0$, $d\theta/d\tau(0)=1$;
    (f) $\epsilon=\pi/8$, $\alpha=0$, $\Gamma=0.244$, $\Delta=0.01$,
    $\lambdat=5\times10^{-4}$, $\theta(0)=0$, $d\theta/d\tau(0)=1$.
  }
  \label{fig:fig5}
\end{figure}

The previous results on parametric resonance are compared to numerical
solutions to (\ref{eq:eqgrav}) in figure \ref{fig:fig5}.  The case of
linear excitation parallel to gravity, $\epsilon=0=\alpha$, is shown
in figure \ref{fig:fig5} (a), where we plot the order $\Delta^{1/2}$
result (\ref{eq:parres-first-sol}), with (\ref{eq:rho1-phi1}), and the
order $\Delta^{3/2}$ result given in (\ref{eq:parres-thrd-sol}) and
Appendix \ref{sec:appa}.  The parameters in that figure were chosen so
$\delta$ is very close to the bound (\ref{eq:secncond}) which in this
case reduces to $\delta=-0.047>-0.05$.  This is the region where
perturbation theory begins to break down, thus appropriate to test the
$\mathcal{O}(\Delta^{3/2})$ correction.  As seen in the figure, the
numerical solution is accurately reproduced by our perturbative
result.  For slightly higher values, $\delta\geq -0.04$, the
$\mathcal{O}(\Delta^{1/2})$ result would be enough to achieve the same
accuracy.  

In figure \ref{fig:fig5} (b) we illustrate the case of linear
excitation not parallel to gravity, $\epsilon=0\neq\alpha$, for which
the order $\Delta^1$ correction (\ref{eq:parres-scnd-sol}), with
(\ref{eq:rho2}), is non-vanishing.  The agreement of the perturbative
results with the numerical one is very accurate.  The decreasing
amplitude of oscillation with increasing $\alpha$ is apparent from the
figure.  As discussed above, (\ref{eq:firstcond}) implies that there
is no resonance when gravity is perpendicular to the linear trajectory
of the suspension point.  This is confirmed in the figure, where the
numerical solution with $\alpha=\pi/2$ is seen to be an oscillating
solution of the type described in section \ref{sec:gravellosc}, with
twice the frequency of the resonant solutions and a much smaller
amplitude.  

The case $\epsilon\neq 0$ is shown in figure \ref{fig:fig5} (c), where
we set $\epsilon=\pi/2$ corresponding to counterclockwise circular
motion of the suspension point. Unlike figure \ref{fig:fig5} (b), the
oscillation amplitude does not show a discernible dependence with
$\alpha$, and for $\alpha=\pi/2$ a resonant solution is obtained.
Aside from those quantitative differences, figures (b) and (c) are
qualitatively similar.  The behaviour for other values of $\epsilon$ is
intermediate between those two.

In figures \ref{fig:fig5} (d), (e) and (f) we show steady-state
rotations with angular frequency 1/2, as described in section
\ref{sec:angfreqhalf}.  It is clear from the figures that the
dependence on $\epsilon$ and $\alpha$ is rather weak.  The lower bound
(\ref{eq:secncond}) is violated in figures (d) and (f), so that a
resonant steady state is not possible in those cases, but it is not
violated in (e), which shows that steady states with angular velocity
1/2 and resonant ones can coexist.  If we had chosen
$d\theta/d\tau(0)$ slightly different from 1 in figure (e), the
solution we would have obtained numerically would have been a resonant
one.  In fact, finding these steady states numerically seems to
require fine tuning of the initial conditions, that being the reason
why we chose the same ones in the three figures.

\section{Final remarks}
\label{sec:finrem}

In the foregoing sections we discussed several steady-state solutions
to the equation of motion (\ref{eq:eqgral}) with parametric
excitation given by (\ref{eq:ellips}).  We
restricted our treatment to the parametrically excited rotator,
($\Delta\lesssim 1$ in (\ref{eq:eqgral})), as opposed to the orbitally
excited rotator ($\Delta\gg 1$), and to low damping,
$10^{-5}\lesssim\lambdat \lesssim 1$.  Similarly, for the pendulum we
considered also $\Gamma\lesssim 1$.  As mentioned in section
\ref{sec:intro}, many other regimes are possible.  Some of those may
be treated as simple variants of the perturbative methods used here,
as occurs, for example, with the two versions of the condition
$\Gamma\lesssim 1$ given in sections \ref{sec:gravellosc}
($\Gamma=\mathcal{O}(\Delta)$) and \ref{sec:gravellosc2}
($\Gamma=\mathcal{O}(\Delta^3)$), which lead to slightly different
perturbative expansions, with qualitatively different results.  Yet,
other parameter regions not considered here may require other
mathematical tools.  But even within the domain of parameter space
considered here, and even within the class of simple steady states
with $\theta(\tau)$ either monotonic or periodic, our results are
certainly not exhaustive.  

In section \ref{sec:stead} we discussed the linearly excited parametric
rotator. We gave the necessary condition for steady-state rotations
and showed that $\theta(\tau)-\tau$ is a periodic function with period
$\pi$ that oscillates about a value $\Theta$ determined by the
parametric amplitude and the damping coefficient.  There is also in
that case a steady state in which the rotator oscillates about the
transverse position.  The latter solution seems to be unstable in the
presence of gravity or eccentricity.  In section \ref{sec:ellipt} we
extended those results to the case of elliptic motion of the suspension
point.  For steady-state rotations we generalised the results for
linear excitation to any eccentricity, including circular excitation,
for direct motion, and established the existence of contrarian steady
state rotations for any elliptical excitation except the circular one.
We treated also steady-state oscillations with the same angular
frequency as the parametric excitation.  Those oscillating
steady-state solutions are characteristic of strictly elliptic
excitation ($0<|\epsilon|<\pi/2$), since they do not exist for linear
or circular motion of the suspension point.  In the linear case the
oscillating solutions reduce to equilibrium ones.  The case of
circular parametric excitation possesses some special characteristics
that cannot be treated with the perturbative method discussed here, so
we will discuss it separately \cite{bouxx}.

In section \ref{sec:gravell} we considered the parametric pendulum.
For steady-state rotations the perturbative theory with
$\Gamma\sim\Delta$ is only a slight modification of that with
$\Gamma=0$ of section \ref{sec:ellipt}, with corrections due to gravity
inducing odd angular frequencies in the spectrum of oscillations about
uniform rotation.  We obtained explicit expressions for those steady
states, as functions of the eccentricity and angle with gravity.  As
in the case without gravity, contrarian-motion solutions exist for all
$\epsilon$ and $\alpha$, except for circular excitation.  The
steady-state oscillatory solutions with $\Gamma\sim\Delta$ are very
different from the case $\Gamma=0$, because $\Theta_0$ in that case is
completely determined by gravity.  By contrast, in the case
$\Gamma\ll\Delta$ the oscillating solutions are a small perturbation
about the case $\Gamma=0$ and, in particular, $\Theta_0$ is
independent of gravity.  This is the essential feature of the
inverted-pendulum phenomenon: because $\Gamma\ll\Delta$, the dynamics
are mainly determined by the torques of the inertial forces due to the
motion of the suspension point, and not by those of gravity.  We
compute the angle $\Theta$ about which these steady states oscillate
as a function of the eccentricity of the suspension point's
trajectory, as measured by the parameter $\epsilon$, and its angle
$\alpha$ with gravity.  That result provides an analytical basis to
the observation that, as a not-yet-inverted pendulum is slowly rotated
from the downward to the upright position, it follows the motion of
the parametric excitation with an angular lag \cite{mic85}.  More
generally, as discussed in quantitative detail in that section and is
intuitively expected, the dependence of the solutions to
(\ref{eq:eqgral}) on the angle $\alpha$ between gravity and the major
axis of the elliptic excitation is stronger for linear parametric
excitation, and grows weaker as $|\epsilon|$ is increased.

In section \ref{sec:parres} we considered parametric resonance, for
elliptic motion of the suspension point with arbitrary eccentricity
and at an arbitrary angle $\alpha$ with gravity.  We found the
explicit form of the oscillatory steady states in the resonance
region, and their dependence with eccentricity and $\alpha$.  Unlike
the usually studied case of linear excitation parallel to gravity,
when the excitation is not linear or not parallel to gravity, integer
angular frequencies enter the spectrum of oscillations, besides the
half-integer frequencies familiar from the case $\epsilon=0=\alpha$.
We obtained necessary conditions for parametric resonance in terms of
$\epsilon$ and $\alpha$.  In particular, for linear excitation the
amplitude for parametric resonance vanishes at $\alpha=\pi/2$, but not
for other eccentricities.  We found also a steady-state rotation mode
with angular frequency equal to half the parametric excitation
frequency, in the parameter region associated with parametric
resonance.  Unlike the resonant solutions, however, the dependence of
the steady-state rotations described in section \ref{sec:angfreqhalf} 
on the parametric excitation amplitude is analytic.  Those rotation
modes  do not seem to have been discussed in the previous literature.

Finally, we want to stress that the steady states discussed above are
only a handful out of a much more numerous spectrum.  For linear and
highly eccentric elliptic excitations there are other modes, even at
small parametric excitation, that are of a rather different nature
than those considered here and will be discussed separately.  For
circular and slightly eccentric elliptic excitations, as mentioned
above, an extension of the methods presented here is necessary to
uncover some of the simplest basic motions \cite{bouxx}.  On the other
hand, for the steady states considered in this paper, the regions of
stability and basins of attraction were probed rather crudely by means
of numerical solutions to the equation of motion.  More detailed and
technically accurate analyses are required in those respects, which
lie beyond the scope of the present paper and which we defer to future
work.

\appendix

\section{Parametric resonance with $\boldsymbol{\epsilon=0=\alpha}$}
\label{sec:appa}

In section \ref{sec:parres} a solution $\theta(\tau)$ at parametric
resonance is given through $\mathcal{O}(\Delta)$.  For
$\epsilon=0=\alpha$, however, the terms of $\mathcal{O}(\Delta)$
vanish, as explained in that section, so the correction at
$\mathcal{O}(\Delta^{3/2})$ becomes important.  In this appendix we
explicitly provide the coefficients $A_3$, $B_3$ in
(\ref{eq:parres-thrd-sol}) in that case.

When $\epsilon=0=\alpha$, consistency of the
$\mathcal{O}(\Delta^{5/2})$ equation for $f_5(\tau)$ requires,
\begin{equation*}
  \label{eq:appaeq}
  \begin{gathered}
    A_3=\frac{N_A}{D}~,
    \qquad
    B_3=\frac{N_B}{D}~,\\
    \begin{aligned}
N_A &= b_1 \cos\varphi_1 + a_1 \sin\varphi_1 + b_3 \cos(3\varphi_1) +
a_3 \sin(3 \varphi_1)~,\\
b_1 &= -16 h\rho_1(256-144\rho_1^2-256 d \rho_1^2 + 3 \rho_1^4)~,\\
a_1 &= \rho_1 (-2^{12} - 2^{13}d-2^{11}\rho_1^2 - 2^9 d\rho_1^2 +
2^{13} d^2 \rho_1^2 + 192 \rho_1^4 - 352 d \rho_1^4 + 3 \rho_1^6)~,\\
b_3 &= 768 h \rho_1^3~,\qquad
a_3 = 48 \rho_1^3~ (16+32 d+\rho_1^2)~,\\
N_B &= b'_1 \cos\varphi_1 + a'_1 \sin\varphi_1 + b'_3 \cos(3\varphi_1) +
a'_3 \sin(3 \varphi_1)~,\\
b'_1 &=  \rho_1 (2^{12} - 2^{13}d-2^{11}\rho_1^2 + 2^9 d\rho_1^2 +
2^{13} d^2 \rho_1^2 - 192 \rho_1^4 - 352 d \rho_1^4 + 3 \rho_1^6)~,\\
a'_1 &= 16 h\rho_1(256+144\rho_1^2-256 d \rho_1^2 + 3 \rho_1^4)~,\\
b'_3 &= 48 \rho_1^3~ (-16+32 d+\rho_1^2)~,\qquad
a'_3 = -768 h \rho_1^3~,\\
D &= 64 (-2^8 + 2^{10} d^2 + 2^8 h^2 -2^7 d \rho_1^2 + 3 \rho_1^4 +
32 \rho_1^2 \cos(2\varphi_1))~,\\
    \end{aligned}
  \end{gathered}
\end{equation*}
with $\rho_1$, $\varphi_1$ as defined in (\ref{eq:rho1-phi1}), and
with $h=\lambdat/\Delta$, $d=\delta/\Delta$.


\begin{thebibliography}{00}
\bibitem{but01} E.\ I.\ Butikov, \emph{``On the dynamic stabilization
    of an inverted pendulum,''} Am.\ J.\ Phys.\ \textbf{69} (2001)
  755--768.
\bibitem{but02} E.\ I.\ Butikov, \emph{``Subharmonic resonances of the
    parametrically driven pendulum,''} J.\ Phys.\ A \textbf{35}
  (2002) 6209--6231.
\bibitem{lan76a} L.\ D.\ Landau, E.\ M.\ Lifshitz,
  \emph{``Mechanics,''} Pergamon, 1976, 80--84. 
\bibitem{arn78a} V.\ I.\ Arnold, \emph{``Mathematical methods of
    classical mechanics,''} Springer-Verlag, 1978, 113--120.
\bibitem{yor78} E.\ D.\ Yorke, \emph{``Square-wave model for a
    pendulum with oscillating suspension,''} Am.\ J.\ Phys.\
  \textbf{46} (1978) 285--288.
\bibitem{cas80} W.\ Case, \emph{``Parametric instability: An
    elementary demonstration and discussion,''}  Am.\ J.\ Phys.\
  \textbf{48} (1980) 218--221.
\bibitem{gor82} I.\ Grosu, D.\ Ursu, \emph{``Simple apparatus for
    obtaining parametric resonance,''} Am.\ J.\ Phys.\ \textbf{50}
  (1982) 325--328.
\bibitem{cur95} F.\ L.\ Curzon, A.\ L.\ H.\ Loke, M.\ E.\
  Lefran\c{c}ois, K.\ E.\ Novik, \emph{``Parametric instability of a
    pendulum,''} Am.\ J.\ Phys.\ \textbf{63} (1995) 132--136.
\bibitem{cay77} T.\ E.\ Cayton, \emph{``The laboratory spring-mass
    oscillator: an example of parametric instability,''}  Am.\ J.\
  Phys.\ \textbf{45} (1977) 723--732.
\bibitem{fal79} L.\ Falk, \emph{``Student experiments on parametric
    resonance,''}  Am.\ J.\ Phys.\ \textbf{47} (1979) 325--328.
\bibitem{ani93} B.\ A.\ Ani\v{c}in, D.\ M.\ Davidovi\'c, V.\ M.\
  Babovi\'c, \emph{``On the linear theory of the elastic pendulum,''}
  Eur.\ Phys.\ J.\ \textbf{14} (1993) 132--135.
\bibitem{yan10} T.\ Yang, B.\ Fang, S.\  Li, W.\ Huang,
  \emph{``Explicit analytical solution of a pendulum with periodically
    varying length,''}  Eur.\ J.\ Phys.\
\textbf{31} (2010) 1089­1096. 
\bibitem{but04} E.\ I.\ Butikov, \emph{``Parametric excitation of a
    linear oscillator,''} Eur.\ J.\ Phys.\ \textbf{25} (2004)
  535--554.
\bibitem{but05} E.\ I.\ Butikov, \emph{``Parametric resonance in a
    linear oscillator at square-wave modulation,''} Eur.\ J.\ Phys.\
  \textbf{26} (2005) 157--174.
\bibitem{row04} D.\ R.\ Rowland, \emph{``Parametric resonance and
    nonlinear string vibrations,''} Am.\ J.\ Phys.\ \textbf{72} (2004)
  758--766. 
\bibitem{ber02} R.\ Berthet, A.\ Petrosyan, B.\ R\c{o}man, \emph{``An
    analog experiment of the parametric instability,''}  Am.\ J.\
  Phys.\ \textbf{70} (2002) 744--749.
\bibitem{fri82} M.\ H.\ Friedman, J.\ E.\ Campana, L.\ Kelner, E.\ H.\
  Seeliger, A.\ L.\ Yergey, \emph{``The inverted pendulum: a
    mechanical analogy of the quadrupole mass filter,''} Am.\ J.\
  Phys.\ \textbf{50} (1982) 924--931.
\bibitem{mic85} M.\ M.\ Michaelis, \emph{``Stroboscopic study of the
    inverted pendulum,''} Am.\ J.\ Phys.\ \textbf{53} (1985)
  1079--1083.
\bibitem{kap65} P.\ L.\ Kapitza, \emph{``Collected Papers,''} D.\ Ter
  Haar editor, Pergamon, 1978, Vol.\ 2, 714--726.
\bibitem{lan76} L.\ D.\ Landau, E.\ M.\ Lifshitz,
  \emph{``Mechanics,''} Pergamon, 1976, 93--95.
\bibitem{arn78} V.\ I.\ Arnold, \emph{``Mathematical methods of
    classical mechanics,''} Springer-Verlag, 1978, 121--122.
\bibitem{nes67} D.\ J.\ Ness, \emph{``Small oscillations of a
    stabilized inverted pendulum,''} Am.\ J.\ Phys.\ \textbf{35}
  (1967) 964--967.
\bibitem{smi92} H.\ J.\ T.\ Smith, J.\ A.\ Blackburn, 
  \emph{``Experimental study of an inverted pendulum,''}  Am.\ J.\
  Phys.\ \textbf{60} (1992) 909--911.
\bibitem{fen98} J.\  G.\  Fenn, D.\  A.\  Bayne, B.\  D.\  Sinclair,
  \emph{``Experimental investigation of the ``effective potential'' of
    an inverted pendulum,''} Am.\ J.\ Phys.\ \textbf{66} (1998)
  981--984.
\bibitem{phe65} F.\ M.\ Phelps, J.\ H.\ Hunter, \emph{``An analytical
    solution of the inverted pendulum,''}  Am.\ J.\ Phys.\ \textbf{33}
  (1965) 285--295; Am.\ J.\ Phys.\ \textbf{34} (1966) 533.
\bibitem{bli65} L.\ Blitzer, \emph{``Inverted pendulum,''}
  Am.\ J.\ Phys.\ \textbf{33} (1965) 1076--1078.
\bibitem{kal70} H.\ P.\ Kalmus, \emph{``The inverted pendulum,''}
  Am.\ J.\ Phys.\ \textbf{38} (1970) 874--878.
\bibitem{pip87} A.\ B.\ Pippard, \emph{``The inverted pendulum,''}
  Eur.\ J.\ Phys.\ \textbf{8} (1987) 203--206.
\bibitem{bla92} J.\ A.\ Blackburn, H.\ J.\ T.\ Smith, N.\
  Groenbech-Jensen, \emph{``Stability and Hopf bifurcations in an
    inverted pendulum,''}  Am.\ J.\ Phys.\ \textbf{60} (1992)
  903--908.
\bibitem{ach95} D.\ J.\ Acheson, \emph{``Multiple-nodding oscillations
    of a driven inverted pendulum,''}  Proc.\ R.\ Soc.\ A \textbf{448}
  (1995) 89­95. 
\bibitem{gra97} W.\  T.\  Grandy, M.\  Schoeck, \emph{``Simulations of
    nonlinear pivot-driven pendula,''}  Am.\ J.\ Phys.\ \textbf{65}
  (1997) 376--381.
\bibitem{mat04} G.\ J.\ Mata, E.\ Pestana, \emph{``Effective
    Hamiltonian and dynamic stability of the inverted pendulum,''}
  Eur.\ J.\ Phys.\ \textbf{25} (2004) 717--721.
\bibitem{rub96} L.\ Ruby, \emph{``Applications of the Mathieu
    equation,''}  Am.\ J.\ Phys.\ \textbf{64} (1996) 39--44.
\bibitem{wolf} S.\ Wolfram, ``The Mathematica Book,''  Third Edition,
  Cambridge U.\ Press, 1996.
\bibitem{wal06} J.\ Walker, \emph{``The flying circus of physics,''}
  Wiley, 2006, 74.
\bibitem{bouxx} A.\ O.\ Bouzas, \emph{``The parametric rotator and
    pendulum with circular excitation,''} in preparation.
\end{thebibliography}
\end{document}